\documentclass{aastex701}
\usepackage{amsmath}
\usepackage{float}
\usepackage{booktabs}
\usepackage{subcaption}
\usepackage{tikz}
\usepackage{tikz-3dplot}
\usepackage{placeins}
\usetikzlibrary{arrows.meta, decorations.markings}

\newcommand{\mbf}[1]{\mathbf{#1}}
\newcommand{\bgt}{\mathbf{B}^>(x, y, 0)}
\newcommand{\blt}{\mathbf{B}^<(x, y, 0)} 
\newcommand{\btor}{\mathbf{B}_T(x, y)}
\usepackage{xcolor}
\renewcommand{\added}[2][]{#2}
\shortauthors{Iyer et al.}

\begin{document}

\title{Validating Coronal Magnetic Field Models Using Gaussian Separation}

\author[orcid=0000-0001-8042-2358]{Abhinav G. Iyer}
\affiliation{Sydney Institute for Astronomy, School of Physics, The University of Sydney, NSW 2006, Australia}
\email[show]{aiye0287@uni.sydney.edu.au}  

\author[orcid=0000-0001-5100-2354]{Michael S. Wheatland} 
\affiliation{Sydney Institute for Astronomy, School of Physics, The University of Sydney, NSW 2006, Australia}
\email{michael.wheatland@sydney.edu.au}

\author[orcid=0000-0003-2244-641X]{Brian T. Welsch} 
\affiliation{Natural \& Applied Sciences, University of Wisconsin-Green Bay, 2420 Nicolet Drive, Green Bay, WI 54311, USA}
\email{welschb@uwgb.edu}

\author[orcid=0000-0002-0671-689X]{Yang Liu}
\affiliation{W.W. Hansen Experimental Physics Laboratory, Stanford University, Stanford, CA 94305-4085, USA}
\email{yliu@sun.stanford.edu}

\author[orcid=0000-0002-3491-077X]{S.A. Gilchrist}
\affiliation{The University of Newcastle, University Dr., Callaghan, NSW 2308, Australia}
\email{stuart.gilchrist@newcastle.edu.au}

\begin{abstract}
Nonlinear Force-free Field (NLFFF) models are widely used to investigate coronal magnetic field structure in solar active regions, but methods to validate them remain limited. Here, we use Gaussian separation, recently applied to solar vector magnetogram data, to assess the accuracy of NLFFF models constructed with two methods: optimization and the current-field iteration (CFIT) implementation of the Grad-Rubin method. Gaussian separation partitions the photospheric vector magnetic field into three components associated with currents flowing below, above, and passing through the photosphere, respectively. Comparing the photospheric field components due to coronal currents in an NLFFF model with those in the original vector magnetogram data provides a check on the accuracy of the model's coronal currents. We consider NLFFF models constructed for the active region AR 11429. The photospheric signatures of coronal currents in both the models and the vector magnetogram data indicate currents flowing above and parallel to central, sheared polarity inversion lines (PILs), consistent with other recent studies. We find that \added{while both models reproduce the coronal current signatures along the upper section of the main PIL, the CFIT model significantly alters the signature of a flux rope along the lower section of the PIL, including shifting its positive-polarity footpoint}. These differences arise from modifications to the vector magnetogram boundary data when solving the NLFFF equations, and from the assumptions underlying the models. We propose Gaussian separation as a useful tool to validate coronal magnetic field models, in addition to existing methods. 
\end{abstract}

\section{Introduction} \label{sec:intro}
    Solar flares and coronal mass ejections (CMEs) are energetic events on the Sun that produce intense radiation, accelerated particles, and launch material into space. Flares and CMEs can produce space weather conditions that disrupt satellites, power grids, communication devices, and can pose risks to astronauts in space \citep[e.g.,][]{tsurutani2012extreme, erinmez2002management, posner2014main}. 
  
    Flares and CMEs are generally understood to be caused by the loss of stability of the coronal magnetic field and the subsequent release of energy stored in coronal electric currents \citep{forbes2000}. Energy accumulates in the corona through magnetic flux emergence and the photospheric shearing of the footpoints of coronal field lines. When the coronal field becomes unstable, part of the accumulated energy is released by magnetic reconnection \citep{priest2002, schrijver2009_flares, sun2012_eruptive}. It is necessary to accurately determine \added{either coronal magnetic fields, or the electric currents that produce the fields}, in order to understand the mechanisms behind flares and CMEs. 

    The most accurate magnetic field determinations, at present, are at the photosphere of the Sun. The vector magnetic field is inferred at the photosphere from spectropolarimetric measurements of magnetically sensitive absorption lines \citep[e.g.,][]{del1996stokes, borrero2011inferring, centeno2014helioseismic}. Full disk vector magnetic field maps, referred to as vector magnetograms, are routinely obtained from the Helioseismic and Magnetic Imager \citep[HMI;][]{hmi} on board the \textit{Solar Dynamics Observatory} \citep[\textit{SDO};][]{sdo} at a spatial sampling of $0.5''$ pixel$^{-1}$  and a cadence of 12 minutes. These data have been used to identify regions of strong magnetic field, known as active regions (ARs), and study their origin and evolution \citep[see, e.g., ][]{centeno2012naked, getling2019origin, wang2021naked}.   

    Photospheric vector magnetogram data are used as boundary conditions for models of the coronal magnetic field. A common \added{technique is to construct} a nonlinear force-free field (NLFFF) model, in which the coronal magnetic field $\mbf{B}$ is assumed static, with zero Lorentz force such that:
        \begin{equation}
            (\nabla \times \mbf{B}) \times \mbf{B} = 0, 
            \label{eq:force_free}
        \end{equation}
    and
        \begin{equation}
            \nabla \cdot \mbf{B} = 0.
            \label{eq:divergence_free}
        \end{equation}
    \added{A variety of methods have been developed to solve Equations \eqref{eq:force_free} and \eqref{eq:divergence_free} \citep[see, e.g.,][]{wiegelmann_review}. Most of the methods use a Cartesian approximation, by neglecting the curvature of the Sun and solving the NLFFF equations in the half-space $z \geq 0$ , with $z=0$ considered as the photospheric boundary. However, methods for solving the NLFFF equations in spherical geometry have also been implemented \citep[see, e.g.,][]{wiegelmann2006_spherical, amari2013_spherical, gilchrist2014_spherical}}. 

    While NLFFF models have been used extensively to reconstruct coronal magnetic fields, it is difficult to validate the results, and different methods have been shown to produce qualitatively and quantitatively different results for the same boundary data \citep[e.g., ][]{schrijver2006, metcalf2008, schrijver2008, derosa2009, derosa_influence_2015}. A basic difficulty with the modeling is that the photospheric vector magnetogram data are generally inconsistent with the force-free assumption. Non-magnetic forces are present in the photosphere \citep[e.g.,][]{metcalf1995force_free, metcalf2008force_free2}. There are also errors in the photospheric vector magnetogram data, due to instrumental noise, resolution limits, and modeling uncertainties due to the inference of the magnetic field from spectropolarimetric data \citep[e.g.,][]{auer1977determination, skumanich1987stokes, bommier2007unnofit}. In addition, there is a $180^{\circ}$ ambiguity in the direction of the field transverse to the line-of-sight, which needs to be resolved in the vector magnetogram data \citep[e.g.,][]{metcalf2006}. 
    
    Since the photospheric vector magnetogram data are generally not force-free, the lower boundary data are often modified during modeling to be more compatible with the NLFFF conditions. The changes to the magnetic field at the lower boundary of the models are introduced either through initial ``preprocessing'' of the boundary data \citep[e.g.,][]{wiegel2006_prepro} and/or while solving the NLFFF equations \citep[see][]{derosa_influence_2015}. In a comparison of different NLFFF methods applied to AR 10978, \citet{derosa_influence_2015} found that the magnitude of the changes in the horizontal magnetic field values at the lower boundary of the models were of the order of several hundred Gauss, significantly exceeding the measurement uncertainties.     
    
    Ideally, NLFFF models for the coronal magnetic field should be validated against observational data to assess their ability to represent the coronal magnetic field. However, there are limited ways to test the models. Previous studies compared NLFFF model magnetic field lines with projected coronal loop structures observed in extreme ultraviolet (EUV) and X-ray images \citep[e.g.,][]{lee1999test, regnier20043d, schrijver2008, derosa2009, sun2012_eruptive}. \citet{derosa2009} found that while some models showed field lines that qualitatively aligned with certain coronal loop structures, the overall match between the model and the observed coronal loops was poor. Additionally, model field lines were also quantitatively compared with stereoscopic determinations of 3D loop paths and found to be substantially different \citep[see][]{derosa2009}. These approaches do not conclusively show whether the models reproduce the coronal magnetic field and currents accurately.

    Recently, \citet{schuck_origin_2022} and \citet{welsch_photospheric_2022} introduced the Gaussian separation theorem \citep[see, e.g.,][]{gauss1877, backus1986, gauss_separation} in the solar context, applying it to photospheric vector magnetogram data to investigate coronal currents. Gaussian separation can be used to partition the photospheric magnetic field into three distinct components, due to currents flowing below, above, or passing through the photospheric boundary, respectively. For simplicity we assume the photosphere is the $x-y$ plane and the three-dimensional (3D) magnetic field is $\mbf{B}(x,y, z)$. The photospheric field $\mbf{B}(x,y, 0)$ can then be written as 
        \begin{equation}
            \mbf{B}(x,y, 0) = \blt + \bgt + \btor + \mbf{B}_M(x,y),
            \label{eq:gs_main_eqn}
        \end{equation}
     where $\bgt$ is the field due to currents above the lower boundary, $\blt$ is the field due to currents below the boundary, $\btor$ is the toroidal component of the photospheric field due to vertical currents passing through the boundary, \added{and $\mbf{B}_M(x,y)$ is any mean field at the photosphere}. The component of the photospheric field $\bgt$ is referred to as the ``photospheric signature'' of the coronal currents. Both \citet{schuck_origin_2022} and \citet{welsch_photospheric_2022} found that the form of $\bgt$ in ARs 10930, 11158 and 12673 implied coronal currents flowing above and along central sheared polarity inversion lines (PILs). \added{In addition, the photospheric signatures of currents passing through the photosphere, $\btor$, allowed the identification of the footpoints of the coronal currents inferred from $\bgt$}.

     \added{These results show the additional insight that can be gained from considering the current $\mathbf{J}$ as the source of the magnetic field, rather than a secondary quantity. The Biot-Savart law attributes the magnetic field at a point $\mathbf{x}$ to a current element at a different point $\mathbf{x}'$, and hence, $\mathbf{B}$ is a non-local function of $\mathbf{J}$. The currents in the corona therefore contribute to the magnetic field at the photosphere. The field $\bgt$ from the Gaussian separation theorem isolates the contribution from the coronal currents, and hence it provides information about the spatial configuration of the coronal currents.}    
     
     In this paper, we investigate the application of Gaussian separation to NLFFF models of the coronal magnetic field. As explained above, NLFFF models have photospheric magnetic field values which differ from the observed vector magnetogram values. Hence, we apply Gaussian separation to the observed vector magnetogram as well as the lower boundary of the NLFFF model. By comparing $\mbf{B}_{\rm obs}^{>}(x,y, 0)$ derived from the observed vector magnetograms with $\mbf{B}^{>}(x,y, 0)$ derived from the lower boundary of the NLFFF model, we can check whether the model accurately represents the coronal currents. We can also compare $\mbf{B}_{T,\text{obs}}(x, y, 0)$ with $\mbf{B}_{T}(x, y)$, to check if the model accurately represents the locations of vertical currents, $J_z(x, y,0)$, crossing the photosphere. 
     
     The structure of the paper is as follows. Section \ref{sec:data_methods} presents the methods being used: Gaussian separation is described in Section \ref{sec:gauss_sep} and the NLFFF methods being applied are  discussed in Section \ref{sec:grad_rubin} and Section \ref{sec:optimization}. Section \ref{sec:ar11429} presents the results for active region AR 11429. The conclusions and discussions are presented in Section \ref{sec:discussion}.

\section{Data and Methods} \label{sec:data_methods}     
    \subsection{Gaussian Separation} \label{sec:gauss_sep}
    Gaussian separation is a technique used to partition the magnetic field on a surface into distinct components, due to current sources above, below or passing through the surface. In its original form, Gaussian separation was used to partition the magnetic field measured on the Earth's surface into two components, either due to sources internal or external to the Earth \citep{gauss1877}. A generalized approach was developed by \citet{backus1986}, to account for sources passing through the surface. Gaussian separation has been used extensively in geomagnetism and planetary physics \citep[see, e.g.,][]{olsen2010}, but was first applied to solar vector magnetogram data by \citet{schuck_origin_2022} in spherical coordinates and \citet{welsch_photospheric_2022} in Cartesian coordinates. In this work, we \added{consider NLFFF models in Cartesian geometry, and hence we use the Cartesian implementation of Gaussian separation described in \citet{welsch_photospheric_2022}}. The photosphere is assumed to be the $z=0$ plane and the 3D magnetic field is $\mbf{B}(x,y,z)$. The photospheric magnetic field, $\mbf{B}(x,y, 0)$ can be partitioned \added{according to Equation \eqref{eq:gs_main_eqn}},   where $\mbf{B}^{<}(x, y, z)$ is the field in $z\geq0$ due to currents $\mbf{J}^<(x, y, z)$ below the photosphere ($z<0$), $\mbf{B}^{>}(x, y, z)$ is the field in $z\leq0$ due to currents $\mbf{J}^>(x, y, z)$ above the photosphere ($z>0$), $\btor$ is the field at $z=0$ due to currents $J_z(x, y, 0)$ passing through the photosphere, \added{and $\mbf{B}_M(x,y)$ is any mean field at the photosphere.}  
    The components $\mbf{B}^{<}(x, y, z)$ and $\mbf{B}^{>}(x, y, z)$, as defined, are potential fields in their respective domains and can be expressed in terms of the gradients of scalar potentials, as $\mbf{B}^{<}(x, y, z) =-\nabla \chi^{<}(x, y, z)$ and $\mbf{B}^{>}(x, y, z)=-\nabla \chi^{>}(x, y, z)$. The components $\mbf{B}^{<}(x, y, z)$ and $\mbf{B}^{>}(x, y, z)$ are each divergence-free and hence the potentials $\chi^{\lessgtr}$ satisfy the 3D Laplace equation, 
            \begin{equation}
                \nabla^2 \chi^{<} = 0, \, z\in [0, \infty), 
                \label{eq:GS_laplace1}
            \end{equation}
            \begin{equation}
                \nabla^2 \chi^{>} = 0, \, z\in (-\infty, 0].
                \label{eq:GS_laplace2}
            \end{equation}
    Since the vertical current density at the photosphere, $J_z(x, y, 0)$ does not produce vertical magnetic fields in the photosphere, $B_z(x, y, 0)$ is entirely due to the currents $\mbf{J}^<(x, y, z)$ and $\mbf{J}^>(x, y, z)$. Hence, 
        \begin{equation}
            B_z(x, y, 0) = -\partial_z \chi^<(x, y, 0)- \partial_z \chi^>(x, y, 0).
            \label{eq:GS_BC1}
        \end{equation}
    Additionally, the horizontal part of the photospheric field, $\mbf{B}_h(x, y, 0)$, produced by the currents $\mbf{J}^<(x, y, z)$ and $\mbf{J}^>(x, y, z)$ has zero curl on $z=0$. This implies that the horizontal divergence of $\mbf{B}_h(x, y, 0)$ satisfies 
        \begin{equation}
            \nabla_h \cdot \mbf{B}_h(x, y, 0) = -\nabla^2_h \chi^<(x, y, 0)- \nabla^2_h \chi^>(x, y, 0). 
            \label{eq:GS_BC2}
        \end{equation}
    Equations \eqref{eq:GS_BC1} and \eqref{eq:GS_BC2} specify two boundary conditions on the unknown potentials $\chi^{\lessgtr}(x, y, 0)$, and each potential satisfies Laplace's equation. This is sufficient to allow determination of $\chi^>(x, y, 0)$ and $\chi^<(x, y, 0)$.

    Following \citet{welsch_photospheric_2022}, the potentials $\chi^{\lessgtr}(x, y, z)$ are expressed in terms of 2D Fourier transforms as 
    \begin{equation}
    \begin{aligned}
        \chi^{\lessgtr}(x, y, z) ={} & \frac{1}{(2 \pi)^2} \int_{-\infty}^{\infty} \mathrm{d}k_x \int_{-\infty}^{\infty} \mathrm{d}k_y \\
        & \times \tilde{\chi}^{\lessgtr}(k_x,k_y) \mathrm{e}^{\mathrm{i}k_xx + \mathrm{i}k_yy \mp k_hz}
    \end{aligned}
    \label{eq:chi_ft}
    \end{equation}
    where the spectral functions $\tilde{\chi}^>(k_x,k_y)$ and $\tilde{\chi}^<(k_x,k_y)$ are the 2D Fourier transforms of $\chi^>(x,y, 0)$ and $\chi^<(x,y, 0)$ respectively, and $k_h = \left|\sqrt{k_x^2 + k_y^2} \right|$. The potential $\chi^{<}(x, y, z) \rightarrow 0$ as $z \rightarrow \infty$, and so each of its Fourier modes decays as $\exp(-k_hz)$ and the potential $\chi^{>}(x, y, z) \rightarrow 0$ as $z \rightarrow -\infty$, and so each of its Fourier modes decays as $\exp(k_hz)$. Equations \eqref{eq:GS_BC1} and \eqref{eq:GS_BC2} can be rewritten using Equation \eqref{eq:chi_ft}, and the spectral functions can be obtained as  
        \begin{equation}
            \tilde{\chi}^{\lessgtr}(k_x, k_y) = \frac{1}{2} \left( \frac{i \mathbf{k}_h \cdot \mathbf{\tilde{B}}_h(k_x, k_y)}{k_h^2} \pm \frac{\tilde{B}_z(k_x, k_y)}{k_h}\right), 
            \label{eq:spectral_fn} 
        \end{equation}
    where $\mathbf{\tilde{B}}_h(k_x, k_y)$ and $\tilde{B}_z(k_x, k_y)$ are the 2D Fourier transforms of $\mathbf{{B}}_h(x, y, 0)$ and ${B}_z(x, y, 0)$ respectively. Equation \eqref{eq:spectral_fn} can then be inverse Fourier transformed to obtain $\chi^>(x,y,0)$ and $\chi^<(x,y,0)$ and hence $\mbf{B}^{<}(x, y, 0)$ and $\mbf{B}^{>}(x, y, 0)$. 

    The field $\btor$ is the toroidal component of $\mathbf{B}(x, y, 0)$ and is produced by vertical currents $J_z(x, y, 0)$ at the photosphere, so it can be written as 
        \begin{equation}
            \btor = \nabla_h \times \left(T(x,y,0)\mbf{\hat{z}}\right),
            \label{eq:btor}
        \end{equation}
    where $T(x,y,0)$ is the toroidal scalar potential. Taking the vertical component of the curl of Equation \eqref{eq:gs_main_eqn} gives
        \begin{equation}
            \mu_0 J_z(x,y, 0) = -\nabla_h^2T(x,y,0),
            \label{eq:tor_pot}
        \end{equation}
    which can be solved using Fourier methods to obtain $T(x,y,0)$ and hence $\btor$. Alternatively, $\btor$ can be obtained by rearranging Equation \eqref{eq:gs_main_eqn}:
        \begin{equation}
            \mbf{B}_{T}(x, y) = \mbf{B}(x, y,0) - \mbf{B}^{<}(x, y, 0) - \mbf{B}^{>}(x, y, 0).
        \end{equation}

    \added{When applying Gaussian separation in a finite Cartesian domain using Fourier methods, the mean field component $\mathbf{B}_M$, needs to be handled explicitly. When $k_h=0$, Equation \eqref{eq:spectral_fn} is singular, and so we explicitly set it to zero during solution. In practice, we compute the domain-averaged horizontal field $\mbf{B}_M = \left<\mathbf{B}_h(x,y,0)\right>$, and subtract it from the observed vector magnetogram data, before applying Gaussian separation. For active region magnetograms with regions of quiet Sun between the strong field and the boundaries, $\mathbf{B}_M$ is typically small and does not significantly affect the decomposition within the region. We note that the spherical version of Gaussian separation \citep[][]{schuck_origin_2022} does not face the above issues, since the harmonic component is zero for a smooth closed manifold such as a sphere.}   
    
    Further details of the Fourier approach to Gaussian separation are given in \citet{welsch_photospheric_2022}. For the Fourier methods in this work, a Python module to obtain the Gaussian separation components is used, and has been made available online \citep[][]{gilchrist_2025_17706429}. \added{In addition, the Carl's Indirect Coronal Current Imager (CICCI) software, described in \citet{schuck_origin_2022}, is also available online, with Cartesian implementations in both IDL and Python} \footnote{https://git.smce.nasa.gov/cicci}.     
    
    \added{Recently, \citet{titov2025} introduced the magnetogram-matching Biot-Savart Law (MBSL), which allows a construction of the coronal magnetic field, such that its radial component ($B_r$) vanishes at the photospheric boundary in spherical geometry. When the MBSL field is superposed with a potential field calculated from the observed $B_r$ distribution, the combined field matches the magnetogram at the lower boundary. However, calculating the potential field from the total normal component at the photosphere, rather than from only the subphotospheric component $\mathbf{B}^<$, misattributes a portion of $B_r$ to fictitious sources solely within the solar interior \citep[e.g., ][]{schuck_origin_2022, schuck2024disentangling}. The MBSL method compensates for this by introducing a set of auxiliary fictitious currents in the solar interior for each current element, either part of a coronal current or a subphotospheric current, such that the combined $B_r$ (of the current element and its associated auxiliary source) at the lower boundary is zero. The MBSL method thus allows a decomposition of the photospheric field into a potential field $\mathbf{B}_0$ calculated from the observed $B_r$, a toroidal field $\mathbf{B}_T$ produced by radial currents $J_r$ and identical to that in the Gaussian decomposition, and a purely tangential poloidal field $\mathbf{B}_{\tilde{S}}$ produced by coronal currents together with their subphotospheric closure currents, and auxiliary fictitious sources. While the coronal current implied by the MBSL method and Gaussian separation may be the same, only the Gaussian decomposition reflects the physical separation of currents by the photosphere, with $\mbf{B}^>$ being free from contributions from subphotospheric currents. Formally, $\mbf{B}_0$ and $\mbf{B}_{\tilde{S}}$ can be expressed as linear combinations of the Gaussian decomposition terms, $\mbf{B}^{<}$ and $\mbf{B}^{>}$. The utility of the MBSL method is in the modeling of coronal magnetic fields, for example, by inserting flux ropes into a potential field without perturbing $B_r$ at the lower boundary.}        

    \subsection{CFIT Method} \label{sec:grad_rubin}
        The first method for constructing an NLFFF which we use here is an implementation of the Grad-Rubin method \citep{grad_rubin} due to \citet{wheatland_calculating_2007}, called current-field iteration (CFIT). The NLFFF equations for the coronal magnetic field $\mbf{B}(x, y, z)$ are written in the alternative form 
            \begin{equation}
                \nabla \times \mbf{B} = \alpha \mbf{B},
                \label{eq:alt_NLFF1}
            \end{equation}
        and 
            \begin{equation}
                (\mbf{B}\cdot\nabla)\alpha = 0, 
                \label{eq:alt_NLFF2}
            \end{equation}
        where $\alpha$ is the force-free parameter. The CFIT method solves Equations \eqref{eq:alt_NLFF1} and \eqref{eq:alt_NLFF2} in a half-space $z>0$ (where $z=0$ is the photospheric boundary), by finding the solution, at iteration $k$, of the linear equations 
            \begin{equation}
                \nabla \times \mbf{B}^{k} = \alpha^{k-1} \mbf{B}^{k-1}
                \label{eq:curlB_alphaB}
            \end{equation} 
            and 
            \begin{equation}
               \left(\mbf{B}^{k} \cdot \nabla \right) \alpha^{k}  = 0.
               \label{eq:B_grad_alpha}
            \end{equation}

        The boundary conditions for the CFIT method are the values of $B_z(x,y,0)$ and the values of $\alpha$ over one polarity of the field at $z=0$, i.e., locations where $B_z(x,y,0)<0$ or $B_z(x,y,0)>0$ respectively. The values of $\alpha$ are required only at one polarity because $\alpha$ is invariant along magnetic field lines, as indicated by Equation \eqref{eq:alt_NLFF2}. The boundary values of $\alpha$ are calculated using 
            \begin{equation}
                \alpha(x, y, 0) = \left. \frac{1}{B_z} \left( \frac{\partial B_y}{\partial x} - \frac{\partial B_x}{\partial y} \right) \right|_{z=0},
                \label{eq:alpha_calculation}
            \end{equation}
    where the magnetic field components are obtained from the vector magnetogram data. \added{Formally, Equation \eqref{eq:alpha_calculation} is valid only for force-free fields, and since the observed photospheric vector magnetogram may not be force-free, it is a source of error for the resulting NLFFF solutions.}     

    The choice of specifying $\alpha$ values over either the positive or the negative polarity of the field results in two possible solutions, which we refer to as the $P$ and $N$ solutions respectively, where $P$ refers to the positive polarity and $N$ refers to the negative polarity. The two solutions will be the same for boundary data which are consistent with the force-free model. However, since the observed photospheric vector magnetograms may be inconsistent with the force-free assumption, the two solutions are different in general \citep[see, e.g., Figure 4 of][]{derosa2009}. 

    The CFIT implementation proceeds as follows. The magnetic field, at a given iteration $k$, is separated into potential and current-carrying components,
        \begin{equation}
            \mbf{B}^{k} = \mbf{B}_0 + \mbf{B}_c^{k},
            \label{eq:pot_nonpot}
        \end{equation}
    where $\mbf{B}_0(x,y,z)$ is the potential field and $\mbf{B}_c^{k}(x,y,z)$ is the current-carrying component of the field. The potential field, $\mbf{B}_0(x,y,z)$, satisfies $\nabla \times \mbf{B}_0 = 0$ and is assumed to satisfy the boundary condition 
        \begin{equation}
            \left. \hat{\textbf{z}} \cdot \textbf{B}_0 \right|_{z=0} = \left. \hat{\textbf{z}} \cdot \textbf{B}_{\text{obs}} \right|_{z=0}.
            \label{eq:pot_component_BC}
        \end{equation}
     Equation \eqref{eq:curlB_alphaB} then implies that the current-carrying component $\textbf{B}_c^{k}$ satisfies $\nabla \times \textbf{B}_c^{k} = \alpha^{k-1} \textbf{B}^{k-1}$. Also, Equation \eqref{eq:pot_component_BC} implies that the current-carrying component satisfies the boundary condition 
        \begin{equation}
            \left. \hat{\textbf{z}} \cdot \textbf{B}_c^{k} \right|_{z=0} = 0. 
            \label{eq:curr_component_BC}
        \end{equation}    

    The current-carrying component is expressed in terms of a magnetic vector potential as $\mbf{B}_c^{k} = \nabla \times \mbf{A}_c^{k}$, and  Equation \eqref{eq:curlB_alphaB} is rewritten as a 3D Poisson equation
        \begin{equation}
            \nabla^2 \mbf{A}_c^{k} = -\alpha^{k-1} \mbf{B}^{k-1}
            \label{eq:vector_poisson}, 
        \end{equation}
        assuming the Coulomb gauge ($\nabla \cdot \textbf{A}_c^{k} =0$). 
        Equation \eqref{eq:vector_poisson} is solved using two-dimensional (2D) Fourier transforms in the $x$ and $y$ directions, and $\mbf{B}_c^{k}(x,y,z)$ is obtained. The details of the solution are described in \citet{wheatland_calculating_2007}.   

        Equation \eqref{eq:B_grad_alpha} is solved using field line tracing. At each point $\mbf{r}$ in the volume, the field line passing through the point is traced in both directions. If the field line connects to $z=0$ at both ends, then $\alpha(\mbf{r})$ is set to the value of $\alpha$ at either the positive or negative polarity of the footpoint of the field line, depending on which solution, $P$ or $N$, is required. If the field line leaves the grid at either the top or side boundaries, $\alpha(\mbf{r})$ is set to zero. This restricts the current in the model to closed field lines.  

        The CFIT method is initiated with a potential field calculated using a Fourier method \citep[][]{allisandrakis1981} and Equations \eqref{eq:curlB_alphaB} and \eqref{eq:B_grad_alpha} are solved iteratively until the values of $\mathbf{B}^k$ and $\alpha^k$ converge at all points in the grid. The convergence of the CFIT method may be assessed by identifying when the normalized magnetic energy, which is defined as the ratio of the total magnetic energy in the volume, $E$, to the energy of the potential field, $E_0$ \citep{wheatland2009_self_consistent} does not change substantially with iteration.

        \added{The CFIT method preserves the lower boundary values of $B_z(x,y,0)$, with iteration, by imposing the boundary condition Equation \eqref{eq:curr_component_BC}. The boundary condition implies the presence of subphotospheric ``mirror'' currents in $z<0$ that close the current path with the coronal currents. This approach is the Cartesian counterpart to the MBSL method \citep[][]{titov2025}, and as explained before, $\mathbf{B}_0$ misattributes a part of the normal component of the field at the lower boundary to fictitious sources solely in the subphotosphere. The mirror currents correct for this and remove the corresponding misattributed current in the subphotosphere. The MBSL method uses spherical geometry, and introduces auxiliary fictitious sources in the solar interior, in addition to the subphotospheric mirror currents. The combined field of the coronal currents, their subphotospheric mirror currents, and the auxiliary fictitious sources has zero radial component at the lower boundary. Hence, $\mathbf{B}_c(x,y,0)$ in the CFIT method is the Cartesian counterpart to $\mathbf{B}_{\tilde{S}} + \mathbf{B}_T$ in the MBSL decomposition.}
        
        The horizontal field values at the lower boundary of the NLFFF solution obtained using the CFIT method are different from the initial vector magnetogram data. This is in part because the boundary values of $\alpha$ due to $B_x(x,y,0)$ and $B_y(x,y,0)$ are used only at one polarity. The $\alpha$ values mapped to the opposite polarity do not necessarily match the vector magnetogram values of $\alpha$. Additionally, points on the grid that are threaded by field lines leaving the grid at the top or side boundaries are assigned $\alpha=0$, and hence boundary points in these open-field regions have $\alpha=0$. These differences in the $\alpha$ values change $B_x(x,y,0)$ and $B_y(x,y,0)$ everywhere in the lower boundary. 

\subsection{Optimization Method} \label{sec:optimization}
    The second method for constructing an NLFFF which we consider is the optimization approach, originally proposed by \citet{wheat_opti} and further developed in \citet{wiegel2004} and \citet{wiegel2010}. The optimization approach is a commonly-used NLFFF method which evolves the magnetic field in the computational volume to minimize a functional, such that if the functional is zero, the field is force- and divergence-free. The functional, $L$, is defined as 
    \begin{equation}
        L = \int_V w_{\text{f}} \frac{\left| (\nabla \times \mbf{B})\times \mbf{B} \right|^2}{|\mbf{B}|^2} + w_{\text{d}} \left| \nabla \cdot \mbf{B} \right|^2 d^3V, 
        \label{eq:opti_functional}
    \end{equation}
     where $\mbf{B}(x, y, z)$ is the coronal magnetic field in the volume $V$, and $w_f$ and $w_d$ are position-dependent weighting functions with values in the range 0 to 1. The lower boundary conditions for the optimization method are all three components of the magnetic field at the photosphere ($z=0$), obtained from the vector magnetogram data. Depending on the approach, these boundary conditions can be used directly, or after ``preprocessing'' to ensure consistency with the force-free assumption, as described below. The method is initialized with a potential field, calculated using a Fourier method \citep{allisandrakis1981}, and is then evolved forward until the functional reaches a minimum stationary value. Since data for the top and side boundaries are not available, the weighting functions are defined in order to reduce the effect of these boundaries on the solution. 
 
     As explained previously, photospheric vector magnetogram data are generally inconsistent with the force-free assumption. When the optimization method is applied to the inconsistent boundary data, the solution field is neither perfectly divergence-free nor force-free. There are two ways to deal with these inconsistencies. The first approach is to ``preprocess'' the observed vector magnetograms such that they are more consistent with necessary but not sufficient conditions for a force-free field \citep{wiegel2006_prepro}. The second approach, described in \citet{wiegel2010}, is to extend the functional in Equation \eqref{eq:opti_functional} by adding a surface integral term,  
        \begin{equation}
            \nu\int_S \left(\mbf{B} - \mbf{B}_{\text{obs}} \right) \cdot \mbf{W} \cdot \left(\mbf{B} - \mbf{B}_{\text{obs}} \right) d^2S, 
            \label{eq:opti_add_functional}
        \end{equation}
      where $\mbf{B}_{\text{obs}}(x,y,0)$ is the observed vector magnetic field at the photosphere ($S$), $\mbf{W}$ is a diagonal matrix containing information about the measurement uncertainties and $\nu$ is the Lagrangian multiplier. In this formulation, differences between the solution field $\mbf{B}(x,y,0)$ at the lower boundary and the observed photospheric field $\mbf{B}_{\text{obs}}(x,y,0)$ are allowed. The elements of the matrix $\mbf{W}$ are chosen such that regions with high measurement uncertainties are allowed to change more significantly. In this study, we use the preprocessing approach. 

      The two approaches to handling the inconsistent boundary data introduce changes to the magnetic field values at the lower boundary in the NLFFF model. The preprocessing approach modifies all three components of the observed magnetic field such that the resulting boundary field is more consistent with the necessary conditions specified in e.g., \citet{molodensky1974equilibrium} and \citet{aly1989reconstruction}. \added{These conditions are integral relations which correspond to the net magnetic force and net torque becoming zero at the lower boundary.} In practice, the observed data are spatially smoothed, leading to a loss of information. In the second approach, since the field in the lower boundary is not constrained during the evolution, all three components of the boundary field are allowed to relax toward more consistent values, resulting in deviations from the vector magnetogram values.
                    
    \subsection{Active Region Selection}
        We consider the active region AR 11429, which produced large X-class flares and is a well-studied region \citep[see, e.g.,][]{elmhamdi2013_ar11429, wang2014, liu2014, sun2015, chintzoglou2015_ar11429}. AR 11429 appeared on the solar disk on 2012 March 4 and was highly flare productive during its disk passage, including producing a powerful X5.4 class flare on 2012 March 7 00:02 UT. The X5.4 flare was also associated with a CME \citep[][]{chintzoglou2015_ar11429, dhakal2020recurring}. We analyze the pre-eruptive configuration of AR 11429 on 2012 March 6 23:36 UT, 26 minutes prior to the X5.4 flare. 
    
    \subsection{Vector Magnetic Field Data}
        The vector magnetic field data used as the lower boundary condition for the NLFFF extrapolations are obtained from the Spaceweather HMI Active Region Patches series \citep[SHARP;][]{sharps}. The SHARP data identifies and tracks active regions on the Sun using line-of-sight magnetograms generated by \textit{SDO}/HMI. In this study, we use the \verb|hmi.sharp_cea_720s| dataset, which provides vector magnetograms of individual active regions, consisting of the field components $B_r$ (the radial component), $B_{\theta}$ (southward component) and $B_{\phi}$ (westward component) on a remapped Cylindrical Equal-Area projection \citep[see, e.g.,][]{coordinates, sun2013coordinate} with a pixel size of $0.5''$ and a cadence of 12 minutes. For the NLFFF codes in this work, the SHARP vector magnetograms are considered on a Cartesian grid, such that $B_x = B_{\phi}$, $B_y = -B_{\theta}$ and $B_z = B_{r}$.   

\section{Analysis of AR 11429} \label{sec:ar11429}
\begin{figure*}
        \centering
        \plotone{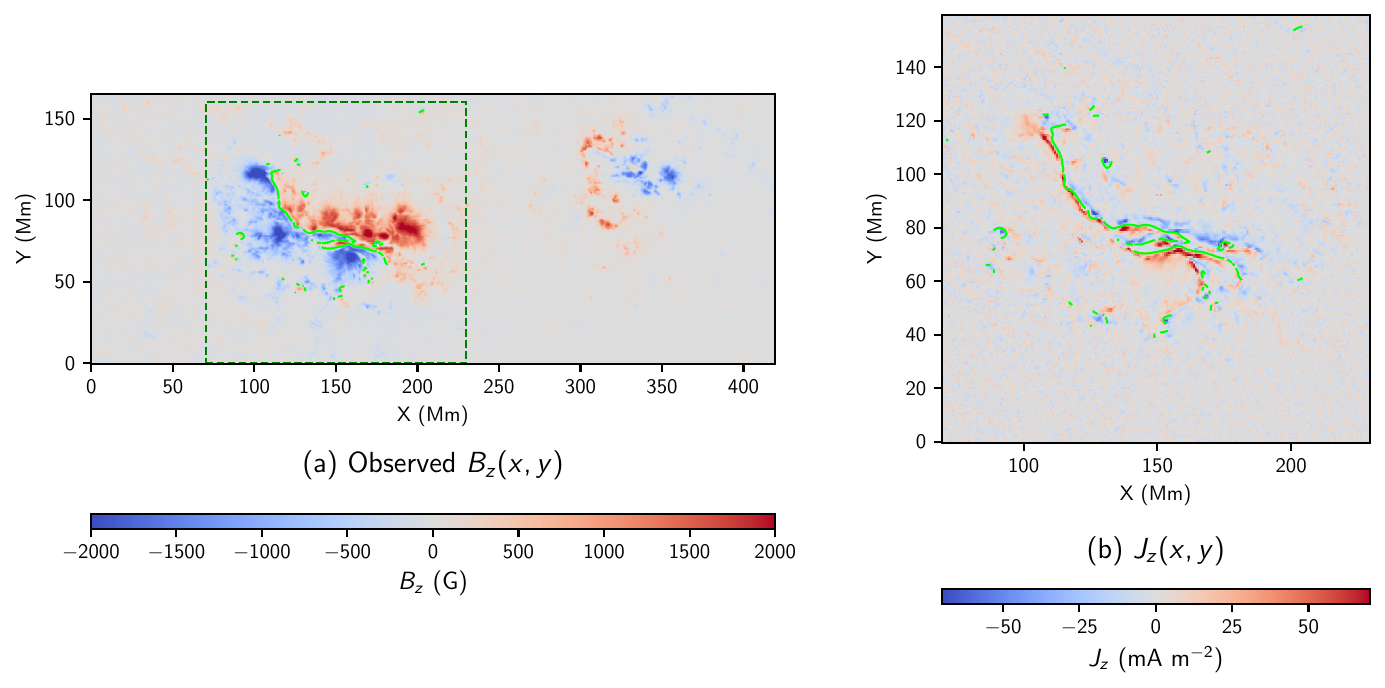}
        \caption{Observations of AR 11429 on 2012 March 6 23:36 UT derived from the SHARP CEA vector magnetogram. (a) Vertical component of the photospheric magnetic field, $B_z(x,y)$, over the SHARP region. The NLFFF models are constructed over the entire field of view and subsequent analysis is carried out in the region within the dashed green box. (b) Vertical current density $J_z(x,y)$ at the photosphere in the region represented by the dashed green box in panel (a). The black $B_z=0$ contours in both panels indicate the main polarity inversion lines (PILs).}
        \label{fig:ar11429_bz+jz_hmi}
    \end{figure*}

    \autoref{fig:ar11429_bz+jz_hmi} shows the vertical magnetic field derived from the SHARP vector magnetogram (panel (a)) and the vertical current density derived from the vertical component of the curl of the magnetic field for AR 11429 on 2012 March 6 at 23:36 UT. The active region exhibits a complex structure, with strong shear along the main polarity inversion line (PIL). Strong vertical currents are also seen along the main PIL, as indicated by panel (b) of \autoref{fig:ar11429_bz+jz_hmi}. 

    \subsection{CFIT Method Applied to AR 11429} \label{sec:cfit_results}
    We construct an NLFFF model of AR 11429 using the CFIT method. The boundary values of $B_z(x, y, 0)$ and $\alpha(x,y, 0)$  for the method are derived from the SHARP vector magnetogram data. Values of $\alpha(x,y, 0)$ are calculated using Equation \eqref{eq:alpha_calculation}, with the derivatives approximated by centered differencing. In weak-field regions, i.e., where $|B_z(x,y,0)| < 0.05 \times \max(|B_z(x,y,0)|)$, we set $\alpha(x,y,0)=0$, since the \added{ratio $J_z / B_z$ may be unreliable} in these regions.

    For the CFIT calculations, we bin the SHARP data over the entire region shown in panel (a) of \autoref{fig:ar11429_bz+jz_hmi} to half resolution resulting in a grid of size $584 \times 230 \times 100$. The region within the dashed green box in panel (a) of \autoref{fig:ar11429_bz+jz_hmi} is considered for subsequent analysis. The grid spacing is $0.72$ Mm pixel$^{-1}$ and is uniform in all three directions. The vertical grid size of 100 pixels is chosen to obtain large-scale magnetic structures, without significantly affecting computational time. 

\begin{figure*}
        \centering
        \plotone{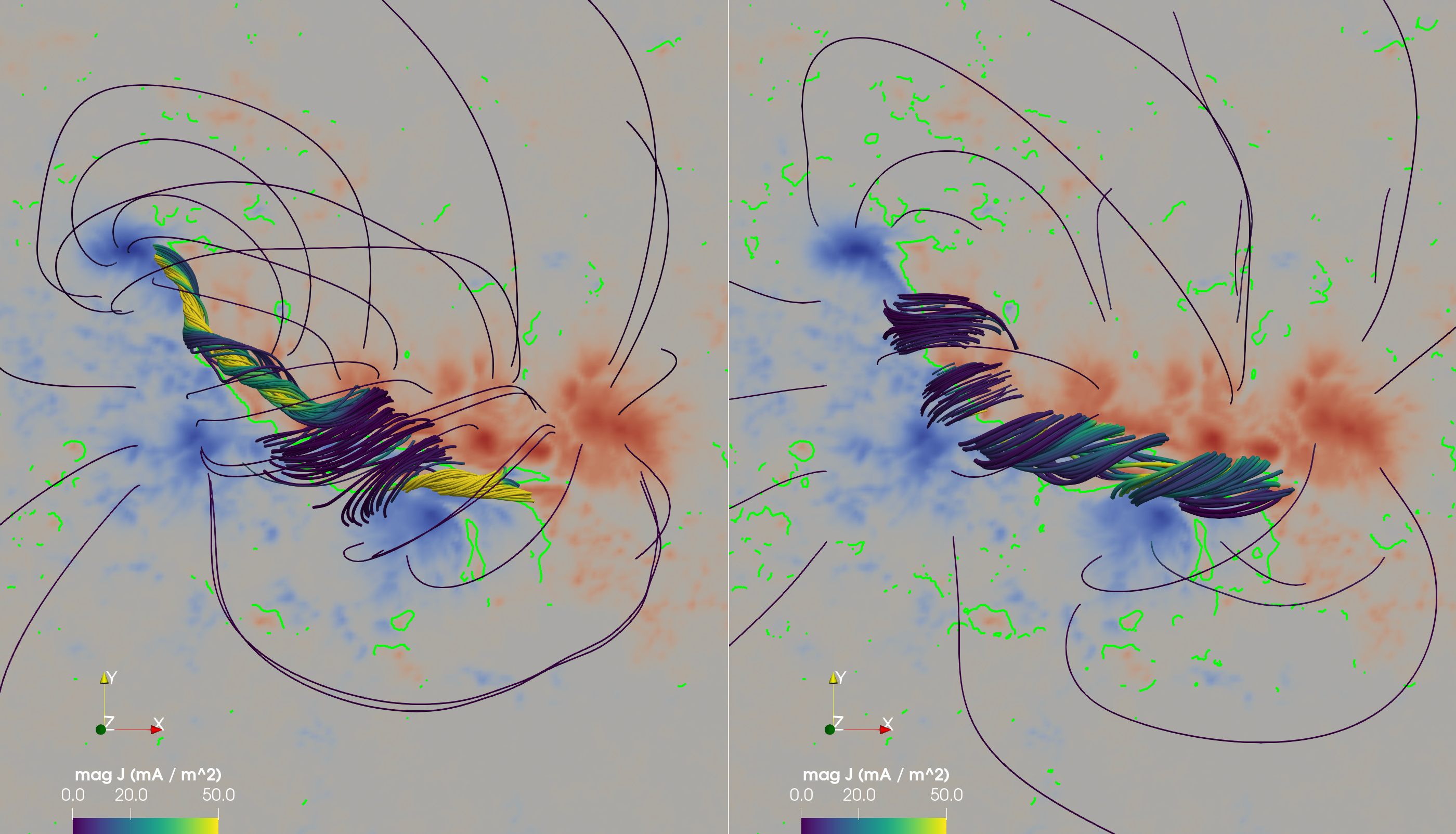}
        \caption{The NLFFF extrapolation of AR 11429 on 2012 March 6 23:36 UT using the CFIT method. The left panel shows the CFIT $N$ solution and the right panel shows the $P$ solution. The background color map denotes the $B_z$ component at the photosphere, with blue denoting $B_z<0$ and red denoting $B_z>0$. The green $B_z=0$ contours indicate the main PILs. The field lines are colored by the magnitude of the current density $\mbf{J}$.}
        \label{fig:field_lines_AR11429_both}
    \end{figure*}

    Selected field lines for the CFIT $N$ (left panel) and $P$ (right panel) solutions are shown in \autoref{fig:field_lines_AR11429_both}. The field lines in each panel are colored by the magnitude of the current density, $|\mbf{J}(x,y,z)|$, sampled along the field line. The background color map indicates $B_z(x,y,0)$ and the green $B_z=0$ contours indicate the main PILs. 

    The CFIT $N$ and $P$ solutions for AR 11429 are significantly different. The $P$ solution has a normalized energy of $E/E_0=1.12$ with field lines generally resembling the initial potential field. Further, the $P$ solution does not reproduce flux ropes along the PIL. The $N$ solution has a normalized energy of $E/E_0 = 1.23$ and includes the flux ropes along the main PIL, similar to results found by previous authors \citep[e.g.,][]{chintzoglou2015_ar11429}. 

    \added{In the $N$ solution (left panel of \autoref{fig:field_lines_AR11429_both}), highly twisted and sheared field lines are observed along the upper-left and lower-right sections of the main PIL, with current density magnitudes of around 50 $\rm{mA}\,\rm{m}^{-2}$ sampled along these field lines, indicating the presence of strong electric currents. Near the middle section of the PIL, the magnitude of the current density along the field lines is significantly lower, indicating a more nearly potential field configuration.}

    \added{In the $P$ solution (right panel of \autoref{fig:field_lines_AR11429_both}), the field lines in the upper-left and middle sections of the PIL are not sheared or twisted, and resemble the potential field configuration. In addition, the current density along the field lines is small in magnitude. Near the lower-right section of the PIL, the field lines are sheared along the PIL, however the current density is smaller in magnitude compared to similar field lines in the $N$ solution.} 
    
\begin{figure*}
        \centering
        \plotone{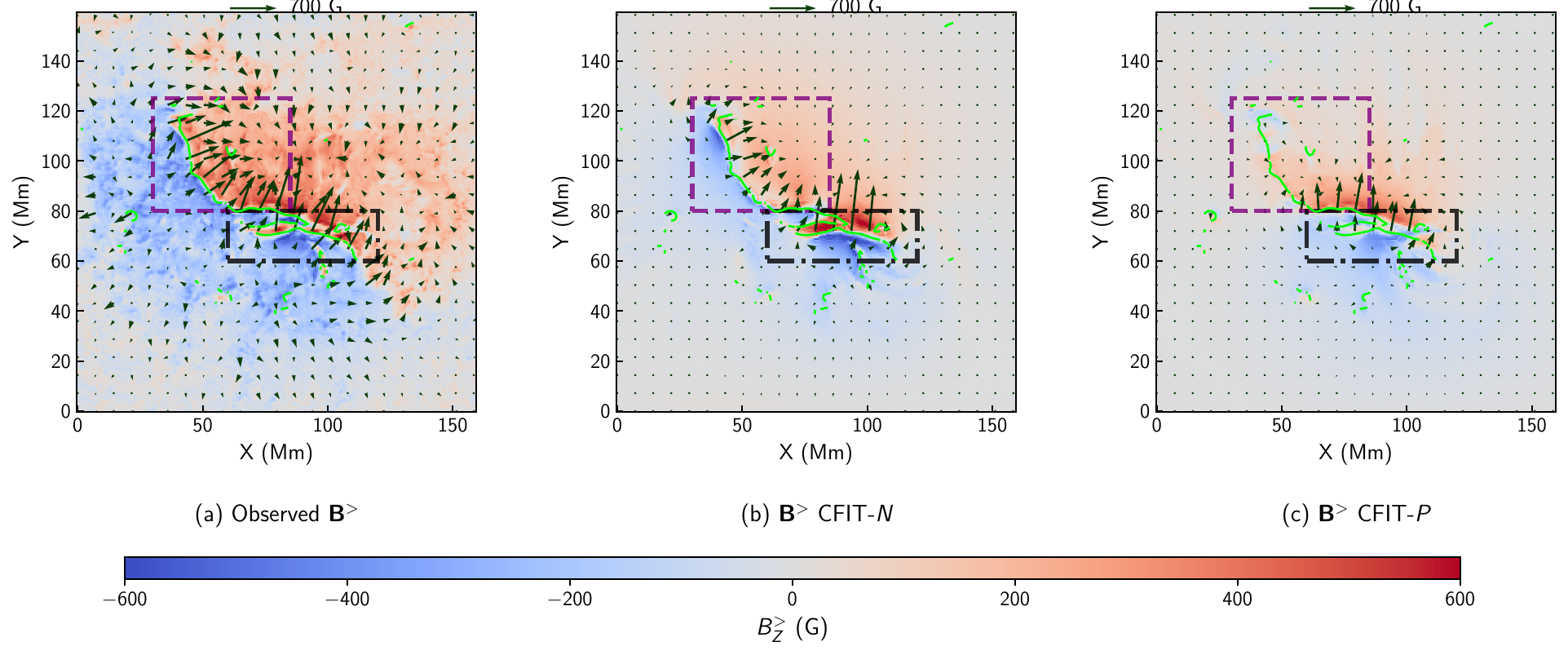}
        \caption{The photospheric signatures of coronal currents, $\bgt$ derived from (a) the SHARP photospheric vector magnetogram, (b) the lower boundary data of the CFIT $N$ solution, and (c) the lower boundary data of the CFIT $P$ solution for AR 11429 on 2012 March 6 23:36 UT. The background color map in each panel shows $B_z^>(x,y,0)$, the black contours show the main PILs corresponding to $B_z(x,y,0) = 0$, and the dark green vectors show the horizontal component $\textbf{B}_h^>(x, y, 0)$. The color and arrow scales are the same for all the  panels. The regions highlighted by the purple and black boxes are discussed in the text.}
        \label{fig:AR11429_B>_comparison}
    \end{figure*}

\begin{figure*}
        \centering
        \plotone{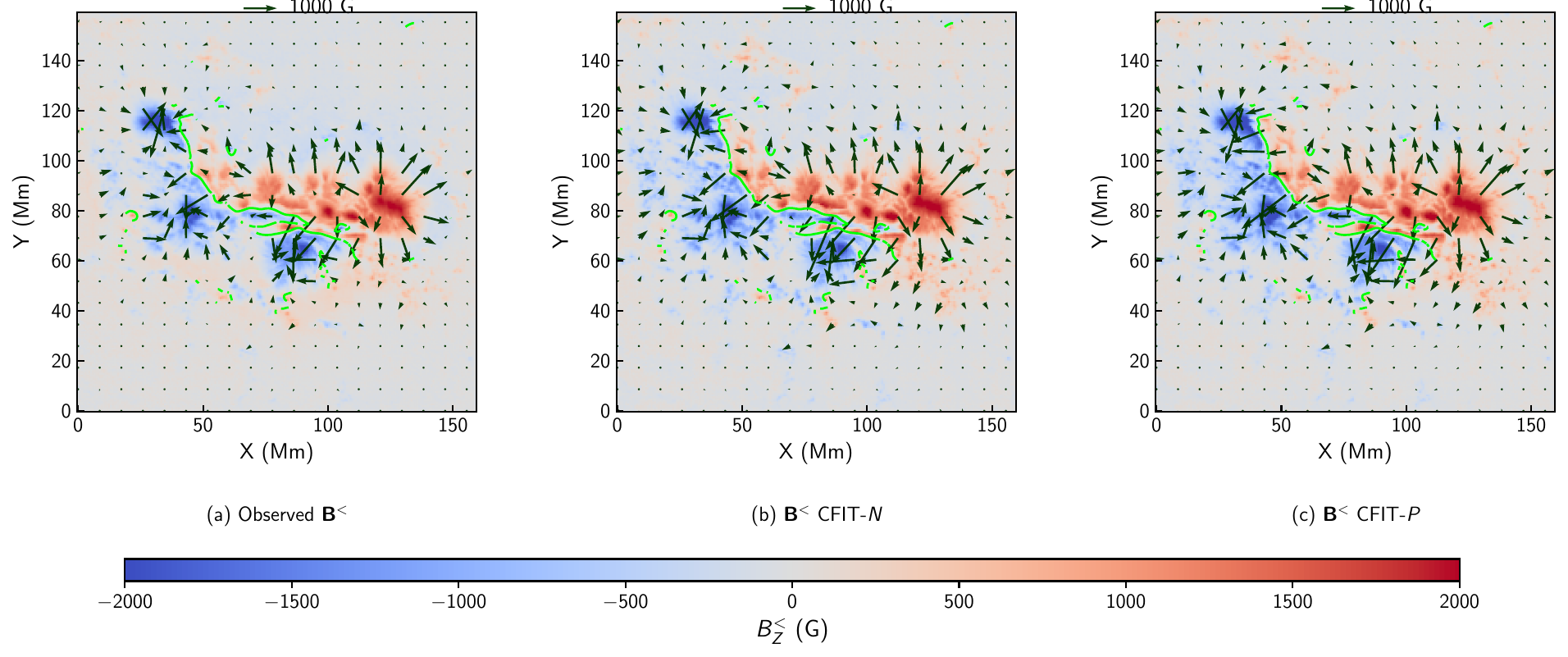}
        \caption{The photospheric signatures of subphotospheric currents, $\blt$ derived from (a) the SHARP photospheric vector magnetogram, (b) the lower boundary data of the CFIT $N$ solution, and (c) the lower boundary data of the CFIT $P$ solution for AR 11429 on 2012 March 6 23:36 UT. The background color map in each panel shows $B_z^<(x,y,0)$, the black contours show the main PILs corresponding to $B_z(x,y,0) = 0$, and the purple vectors show the horizontal component $\textbf{B}_h^<(x, y,0)$.}
        \label{fig:AR11429_B<_comparison}
    \end{figure*}

\begin{figure*}
    \centering
    \plotone{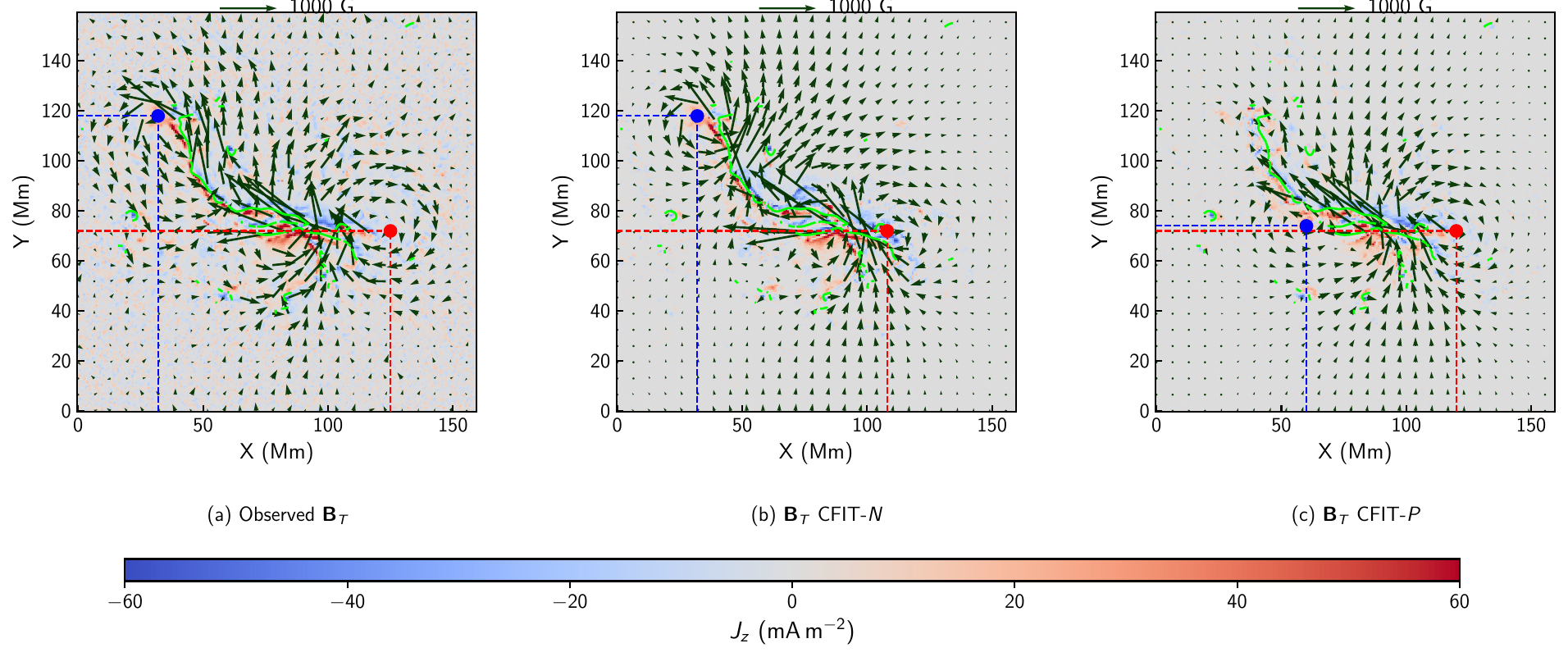}
    \caption{The toroidal field, ${\textbf{B}_{T}(x,y)}$,  produced by the vertical currents, $J_z(x,y, 0)$ derived from (a) the SHARP photospheric vector magnetogram, (b) the lower boundary data of the CFIT $N$ solution, and (c) the lower boundary data of the CFIT $P$ solution for AR 11429 on 2012 March 6 23:36 UT. The background color map in each panel shows $J_z(x,y)$, the black contours show the main PILs corresponding to $B_z(x,y)=0$, and the orange vectors show $\btor$.}
    \label{fig:AR11429_B_T_comparison}
\end{figure*}

    Figures \ref{fig:AR11429_B>_comparison}-\ref{fig:AR11429_B_T_comparison} show the result of applying Gaussian separation to the SHARP photospheric vector magnetogram and to the lower boundary data of the \added{CFIT $N$ and $P$ solutions} for AR 11429. \autoref{fig:AR11429_B>_comparison} shows the component of the photospheric field produced by coronal currents, $\bgt$ derived from the SHARP vector magnetogram (panel (a)), the lower boundary of the CFIT $N$ solution (panel (b)), \added{and the lower boundary of the CFIT $P$ solution (panel (c)).} The background color map in each panel denotes the vertical component, $B_z^>(x,y, 0)$, and the dark green vectors denote the horizontal component $\textbf{B}_h^>(x, y,0) = (B_x^>, B_y^>)$. The direction of the $\textbf{B}_h^>(x, y, 0)$ vectors in panel (a) is consistent with a large-scale coronal current flowing from the upper-left of the AR, at around $(x,y)=(40 \, \text{Mm}, 120\, \text{Mm})$, to the lower-right, at around $(x,y)=(110 \, \text{Mm}, 60\, \text{Mm})$, above and along the PIL. In panel (b), the $\textbf{B}_h^>(x, y,0)$ vectors along the PIL in the region highlighted by the dashed purple box are qualitatively similar to those in panel (a). \added{In panel (c), there are no vectors along the PIL within the purple box, indicating a significant deviation from the vector magnetogram}. There are specific differences between the results --- for example, the vectors along the PIL within the black box in panels (b) and (c) are not oriented in the same direction as those in panel (a). The discrepancies may be attributed to the changes in the boundary values of $B_x(x,y, 0)$ and $B_y(x,y, 0)$ that occur during the construction of the NLFFF solution. The changes in the horizontal field affect the inferred $\bgt$ and $\blt$, as discussed below. In addition, $\bgt$ is significantly weaker in the weak-field regions of both the CFIT solutions due to the process of setting $\alpha(x,y, 0) = 0$ in open field regions, which removes the currents in these regions. Despite the differences, there is qualitative correspondence between $\bgt$ derived from the CFIT $N$ solution and the vector magnetogram.  

    To understand how changes in the boundary values of  $\mathbf{B}_h(x,y,0)$ affect the differences between $\mbf{B}^{\lessgtr}(x,y, 0)$ derived from the vector magnetogram and the lower boundary of the CFIT solution, we consider Equation \eqref{eq:spectral_fn}. Denoting the difference between the spectral functions from the vector magnetogram and the model as $\Delta \tilde{\chi}^{\lessgtr}_{\mathrm{CFIT}} = \tilde{\chi}^{\lessgtr}_{\mathrm{obs}} - \tilde{\chi}^{\lessgtr}_{\mathrm{CFIT}}$, we have 
    \begin{equation}
    \begin{split}
    \Delta \tilde{\chi}^{\lessgtr}_{\mathrm{CFIT}} = \frac{1}{2} \frac{i \mathbf{k}_h}{k_h^2} \cdot \left[\mathbf{\tilde{B}}_{h,\mathrm{obs}}(k_x, k_y) - \tilde{\mathbf{B}}_{h,\mathrm{CFIT}}(k_x, k_y)\right] \\ 
    \mp \frac{\tilde{B}_{z,\mathrm{obs}}(k_x, k_y) - \tilde{B}_{z, \mathrm{CFIT}}(k_x, k_y)}{k_h}, 
    \label{eq:theoretical_diff_b_gt_lt}
    \end{split}
    \end{equation}
 where $\mathbf{\tilde{B}}_{h,\mathrm{obs}}(k_x, k_y)$ and $\tilde{B}_{z,\mathrm{obs}}(k_x, k_y)$ are derived from the  vector magnetogram and $\tilde{\mbf{B}}_{h,\mathrm{CFIT}}(k_x, k_y)$ and $\tilde{B}_{z,\mathrm{CFIT}}(k_x, k_y)$ are derived from the lower boundary of the CFIT solution. As mentioned previously, the CFIT method preserves $B_z(x,y,0)$ at the lower boundary by construction and hence $\tilde{B}_{z,\mathrm{obs}}(k_x, k_y) = \tilde{B}_{z,\mathrm{CFIT}}(k_x, k_y)$. Equation \eqref{eq:theoretical_diff_b_gt_lt} then becomes 
    \begin{equation}
    \Delta \tilde{\chi}^{\lessgtr}_{\mathrm{CFIT}} = \frac{1}{2} \frac{i \mathbf{k}_h}{k_h^2} \cdot \left[\mathbf{\tilde{B}}_{h,\mathrm{obs}}(k_x, k_y) - \tilde{\mathbf{B}}_{h,\mathrm{CFIT}}(k_x, k_y)\right], 
    \end{equation}
 which implies that differences in $\bgt$ and $\blt$ between the vector magnetogram and the model are due entirely to the differences in the total horizontal component $\mathbf{B}_h(x,y, 0)$. Further, \added{the magnitude of the differences in $\bgt$ and $\blt$ will be equal.} 

 \added{Following the study of NLFFF models in \citet{derosa_influence_2015}, we quantify the differences in $\bgt$, $\blt$, and $\btor$ using the change in the vector field at $z=0$, defined as  
        \begin{equation}
            \mathbf{\Delta}\mathbf{B}(x,y,0) = \mbf{B}_{\mathrm{obs}}(x,y,0) - \mathbf{B}_{\mathrm{CFIT}}(x,y,0),
            \label{eq:change_in_vec_field}
        \end{equation}
 where $\mbf{B}_{\mathrm{obs}}(x,y,0)$ is derived from the input vector magnetogram and $\mathbf{B}_{\mathrm{CFIT}}(x,y,0)$ is from the lower boundary of the CFIT $N$ or $P$ solution. We consider the magnitude of the differences in the horizontal components, $|\Delta\mathbf{B}_h| = \sqrt{(\Delta B_x)^2 + (\Delta B_y)^2 }$, and the vertical components, $|\Delta B_z|$ separately. The rms values for the magnitude of the differences are evaluated for the total field and the Gaussian separation components, within the region highlighted by the green box in panel (a) of \autoref{fig:ar11429_bz+jz_hmi}. Table~\ref{tab:diff_bgt_blt_CFIT} lists the above rms values of the differences for the CFIT $N$ and $P$ solutions, and the optimization solution discussed in Section \ref{sec:opti_results}}. 

\begin{table}[ht]
    \centering
    \caption{Differences (rms) between vector magnetogram and the lower boundary data for CFIT $N$, $P$, and optimization solutions.}
    \label{tab:diff_bgt_blt_CFIT}
    \begin{tabular}{lcccc}
        \toprule
        & \multicolumn{4}{c}{rms [G]} \\
        \cmidrule(lr){2-5}
        Method & $\Delta \mathbf{B}_h / \Delta B_z$ & $\Delta \mathbf{B}_h^{>} / \Delta B_z^{>}$ & $\Delta \mathbf{B}_h^{<} / \Delta B_z^{<}$ & $\Delta \mathbf{B}_T$ \\
        \midrule
        CFIT $N$ & 204.7 / \nodata & 83.6 / 80.6 & 83.6 / 80.6 & 116.6 \\
        CFIT $P$ & 214.7 / \nodata & 95.9 / 93.4 & 95.9 / 93.4 & 109.3 \\
        \midrule
        Optimization & 84.2 / 53.0 & 37.0 / 33.1 & 52.4 / 46.9 & 59.9 \\
        \bottomrule
    \end{tabular}
\end{table}

\added{The CFIT method preserves $B_z(x,y,0)$ at the lower boundary, and hence $|\Delta B_z|$ vanishes. However, $B_z^<$ and $B_z^>$ are not similarly preserved, leading to non-zero values in $\Delta B_z^{\lessgtr}$. Additionally, as explained previously, the RMS differences in $\bgt$ and $\blt$ are equal, for both the CFIT solutions.}  

\added{\autoref{fig:AR11429_B_T_comparison} compares the toroidal field, $\mathbf{B}_T(x,y)$ produced by the vertical currents, $J_z(x,y, 0)$ passing through the photosphere. Panel (a) shows $\btor$ derived from the SHARP vector magnetogram, while panels (b) and (c) show $\btor$ derived from the lower boundary data of the CFIT $N$ and $P$ solutions,  respectively.  The background color map in each panel shows $J_z(x,y, 0)$ and the dark green vectors show $\btor$. The vortices in $\mbf{B}_T(x,y)$, marked by the blue and red circles in panel (a) indicate the regions where the vertical currents are directed out of (counterclockwise vortex) and into (clockwise vortex) the photosphere,  respectively. These oppositely directed currents together contribute a substantial fraction of the sheared field along the PIL. The vortices are also observed in panel (b) for the $N$ solution, but the location of the clockwise vortex (red circle) is different. This discrepancy is because $\alpha(x,y, 0)$ values mapped from the negative polarity to the positive polarity in the $N$ solution do not match the vector magnetogram values of $\alpha(x,y, 0)$ in the positive polarity. This implies that the modeled vertical currents may not be at the correct locations as indicated by the values of $\btor$ derived from the vector magnetogram. In the $P$ solution (panel (c)), the counterclockwise vortex (blue circle) is quite far from the location seen in panel (a). However, the location of the clockwise vortex (red circle) is much closer to that observed in panel (a). Since the $P$ solution uses $\alpha(x,y,0)$ values from the positive polarity, the vertical currents inferred from $\btor$ are more consistent with those inferred from the vector magnetogram in the positive polarity.}  

We now consider the currents, $\mbf{J}_{\mathrm{CFIT}}(x,y, z)$, in the CFIT $N$ solution in order to identify the structures that produce $\bgt$ at the lower boundary. We consider the $N$ solution for further analysis here since it is more highly non-potential and more consistent with previous studies. The currents are calculated using Ampère's law as $\mu_0\mbf{J}_{\mathrm{CFIT}}(x, y, z) = \nabla \times \mbf{B}_{\mathrm{CFIT}}(x,y,z)$, where $\mbf{B}_{\mathrm{CFIT}}(x,y,z)$ is the magnetic field in the NLFFF solution, and the derivatives are approximated using centered differencing. 
\begin{figure*}
    \plotone{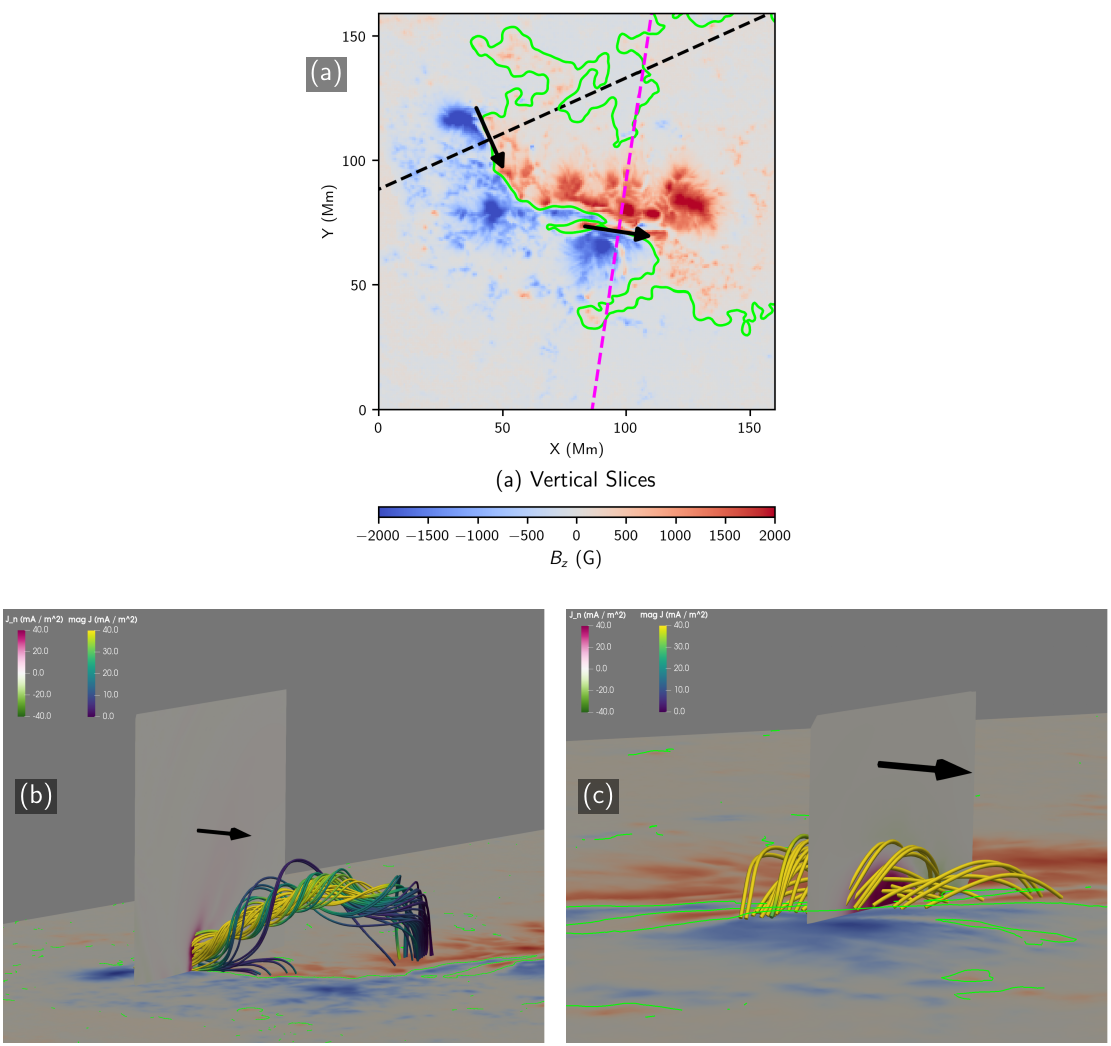}
    \caption{Flux rope and associated current density distribution for AR 11429, derived from the CFIT $N$ solution. (a) Dashed lines indicate the locations of vertical slices perpendicular to the main PIL considered for analysis. The lower boundary shows $B_z(x,y, 0)$ and the main PIL, similar to previous figures. (b) Current density in a vertical slice perpendicular to the upper section of the PIL (black dashed line in panel (a)). The field lines passing through the vertical slice are colored by $|\mbf{J}_{\mathrm{CFIT}}(x,y, z)|$, sampled along the line. The normal current density in the slice, $J_n$ is positive for currents flowing in the direction of the normal vector to the plane. (c) Similar to panel (b) but for a vertical slice perpendicular to the lower section of the PIL (magenta dashed line in panel (a)).}
    \label{fig:AR11429_Upper_Lower_PIL_CFIT}
\end{figure*}

\added{\autoref{fig:AR11429_Upper_Lower_PIL_CFIT} shows the flux rope along the upper-left (left panel) and lower-right (right panel) sections of the PIL, and the associated current density distributions, $J_n$, in vertical cross-sections perpendicular to the PIL. The field lines are colored by the magnitude of the current density sampled along the line. The color map in the vertical cross-section is such that currents flowing along the normal vector (black arrow) are positive. In the upper-left section of the PIL, the peak value of $|J_n| = 47.03$ mA m$^{-2}$, is at around $z = 3.6$ Mm, indicating a relatively low-lying flux rope \citep[see][for a similar result]{chintzoglou2015_ar11429}. Comparing the $J_n$ distribution in \autoref{fig:AR11429_Upper_Lower_PIL_CFIT} with $\bgt$ in the purple box of panel (b) in \autoref{fig:AR11429_B>_comparison} suggests that the $\bgt$ structure along the PIL primarily arises from the strong low-lying flux rope.}

\added{In the lower-right section of the PIL, the peak current in the vertical cross-section is $|J_n| = 57.08$ mA m$^{-2}$ and occurs at around $z=0.72$ Mm, indicating a stronger and lower-lying flux rope compared to the flux rope along the upper-left half of the PIL. We again compare the $J_n$ distribution in the right panel of \autoref{fig:AR11429_Upper_Lower_PIL_CFIT} with $\bgt$ in the blue box of panel (b) in \autoref{fig:AR11429_B>_comparison}, and infer that the primary contribution to $\bgt$ is from the strong, low-lying flux rope. As discussed previously, the structure of $\bgt$ derived from the CFIT $N$ solution lower boundary in the blue box (panel (b) of \autoref{fig:AR11429_B>_comparison}) is different from $\bgt$ derived from the vector magnetogram. This indicates that the field lines in the lower-right section of the PIL may not be modeled accurately.}

\begin{figure*}
    \centering
    \plotone{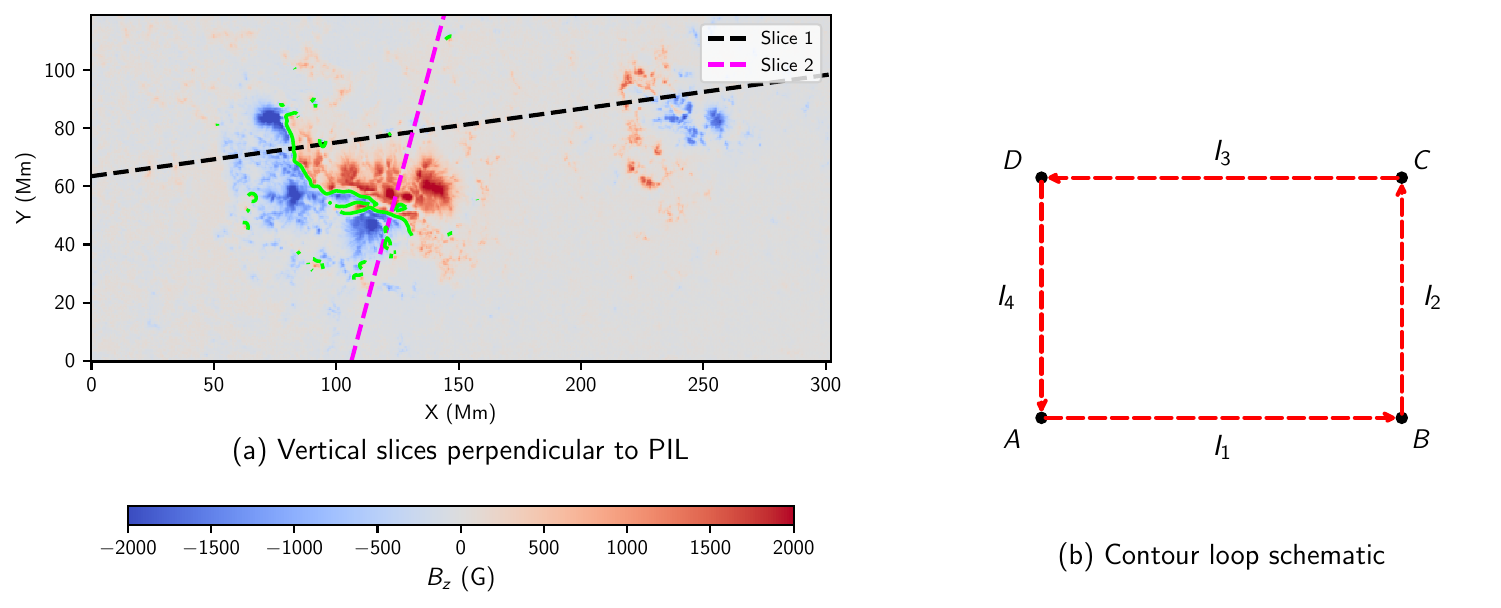}
    \caption{(a) The dashed lines, perpendicular to the PIL, show the baseline for the vertical cross-sections over which Ampère's law is applied. (b) The schematic of the contour loop indicating the integrals considered in the text.} 
    \label{fig:iphot_icor_cfit}
\end{figure*}

To quantitatively assess the magnitude of the coronal currents in the NLFFF model, we consider a metric involving Ampère's law, 
    \begin{equation}
        \oint_C \mathbf{B} \cdot d\mbf{l} = \mu_0 \iint_S \mathbf{J} \cdot d\mathbf{S},
        \label{eq:amp_law}
    \end{equation}
where the magnetic field $\mathbf{B}(x,y,z)$ is integrated around a closed contour $C$, and the total current is evaluated on the surface $S$ enclosed by $C$. \added{We denote the contour integral on the left-hand side of Equation \eqref{eq:amp_law} as $I_C$ and the surface integral on the right-hand side as $I_S$. We consider two vertical cross-sections perpendicular to the PIL, as illustrated in panel (a) of \autoref{fig:iphot_icor_cfit}, and apply Equation \eqref{eq:amp_law}. The left-hand side of Equation \eqref{eq:amp_law} can be decomposed as 
    \begin{equation}
        I_C = I_1 + I_2 +I_3+I_4, 
        \label{eq:ic_decomp}
    \end{equation}
where the integrals $I_1$ to $I_4$ are along the sections indicated in panel (b) of \autoref{fig:iphot_icor_cfit}. The individual integrals along each section of the contour, along with $I_S$ are evaluated for the CFIT solution and listed in \autoref{tab:integrals_cfit}.}
\begin{table}[ht]
    \centering
    \caption{Ampère's law integrals
    for the vertical cross-sections of the CFIT $N$ solution indicated in \autoref{fig:iphot_icor_cfit}.}
    \begin{tabular}{lcc}
        \toprule
        Quantity & Slice 1 ($10^9$\,A) & Slice 2 ($10^9$\,A)\\
        \midrule
        $I_1$ (observed)   & 1915.5  & 3076.8 \\
        $I_1$ (CFIT)       & 2510.8  & 2411.2 \\
        $I_2$              & $-38.4$ &   29.7 \\
        $I_3$              &  $-1.9$ &   83.9 \\
        $I_4$              &   35.1  &   82.2 \\
        \midrule
        $I_C$       & 2505.55  & 2607.04 \\
        $I_S$ & 2476.74  & 2625.31 \\
        \bottomrule
    \end{tabular}
    \label{tab:integrals_cfit}
\end{table}

\added{In both cross-sections, $I_C \approx I_S$, which is expected, since the CFIT method is based on Ampère's law. The primary contribution to $I_C$ is from the integral along the lower boundary, and hence $I_1$ may be used as a metric for the net current through the cross-section. When $I_1$ is evaluated using $\mbf{B}_{\mathrm{obs}}(x,y,0)$, instead of $\mbf{B}_{\mathrm{CFIT}}(x,y,0)$, there are significant differences. Along a line perpendicular to the upper section of the PIL (Slice 1 in panel (a) of \autoref{fig:iphot_icor_cfit}) the observed value of $I_1$ is around $31 \%$ lower than that calculated using the CFIT $N$ solution. However, in the lower section of the PIL (Slice 2 in panel (a) of \autoref{fig:iphot_icor_cfit}), the observed value of $I_1$ is around $15\%$ higher. The discrepancies in the inferred currents are primarily due to the modifications to the lower boundary data made during the CFIT calculation.}

\added{We note that the line integral of $\bgt$ on the photosphere does not provide an estimate of the  coronal currents crossing the plane above the photosphere.} As explained in Section \ref{sec:gauss_sep}, the components $\bgt = -\nabla \chi^>(x,y, 0) $ and $\blt = -\nabla \chi^<(x,y, 0)$ are both potential fields at $z=0$. The line integral of $\mathbf{B}^{\lessgtr}(x,y,0)$ between two points $\mathbf{r}_1$ and $\mathbf{r}_2$ is then  
    \begin{equation}
    \begin{split}
        \int_{\mathbf{r}_1}^{\mathbf{r}_2} \mathbf{B}^{\lessgtr}(x,y,0) \cdot d \mathbf{l} &= -\int_{\mathbf{r}_1}^{\mathbf{r}_2} \nabla \chi^{\lessgtr}(x,y,0) \cdot d \mathbf{l} \\
             &= \chi^{\lessgtr}(\mathbf{r}_1) - \chi^{\lessgtr}(\mathbf{r}_2).  
    \end{split}
    \end{equation}
\added{Hence, the line integral depends only on the values of $\chi^{\lessgtr}(x,y,0)$ at the endpoints, and vanishes when $\chi^{\lessgtr}(\mathbf{r}_1) = \chi^{\lessgtr}(\mathbf{r}_2)$. In general, if the integration path is extended to infinity, then we expect that the line integral of $\mathbf{B}^{\lessgtr}(x,y,0)$ will be zero, because the values of $\chi^{\lessgtr}(x,y, 0)$ approach a constant at infinity.
For a finite integration path (e.g.\ across a magnetogram), the values of $\chi^{\lessgtr}(x,y,0)$ at the endpoints may not strictly be equal, and so the line integral of $\mathbf{B}^{\lessgtr}(x,y,0)$ will not be exactly zero. } 

\added{Since the line integrals of $\blt$ and $\bgt$ are zero along the lower boundary, it follows that the toroidal component, $\btor$, provides the sole contribution to the line integral of the total field $\mbf{B}(x,y,0)$. This is somewhat counterintuitive: $\btor$ is defined completely by $J_z(x,y,0)$ at the photosphere, yet it provides information on the coronal currents crossing the plane above. We note, however, that the models require $\nabla \cdot \mbf{J} = 0$, and so $J_z(x,y,0)$ values at the photosphere imply currents which must close in the corona above. In \autoref{sec:appendix}, we show the contributions of $\blt$, $\bgt$, and $\btor$ to the line integral of the total field in a plane for a simple analytic test case: a ring current with axis in the plane. We confirm that only $\btor$ contributes to the line integral of the total field, and that the contributions from $\blt$ and $\bgt$ vanish. We defer further investigation of the relation between the line integrals of Gaussian separation components and currents in the corona to a future study.}

\subsection{Optimization Method Applied to AR 11429} \label{sec:opti_results}
    We have also constructed an NLFFF model of AR 11429 using the optimization method. The boundary conditions for the optimization method are obtained after preprocessing the vector magnetogram data, following the method of \citet{wiegel2006_prepro}. The preprocessing step involves four weighting parameters and the values selected here are $\mu_1 = \mu_2 =1$ and $\mu_3 = \mu_4=10^{-3}$ \citep[see][for a description of the weighting parameters]{wiegel2006_prepro}. We use a grid of size $583 \times 230 \times 100$ for the optimization calculations to match the grid used for the CFIT method. 

\begin{figure*}
            \centering
            \plotone{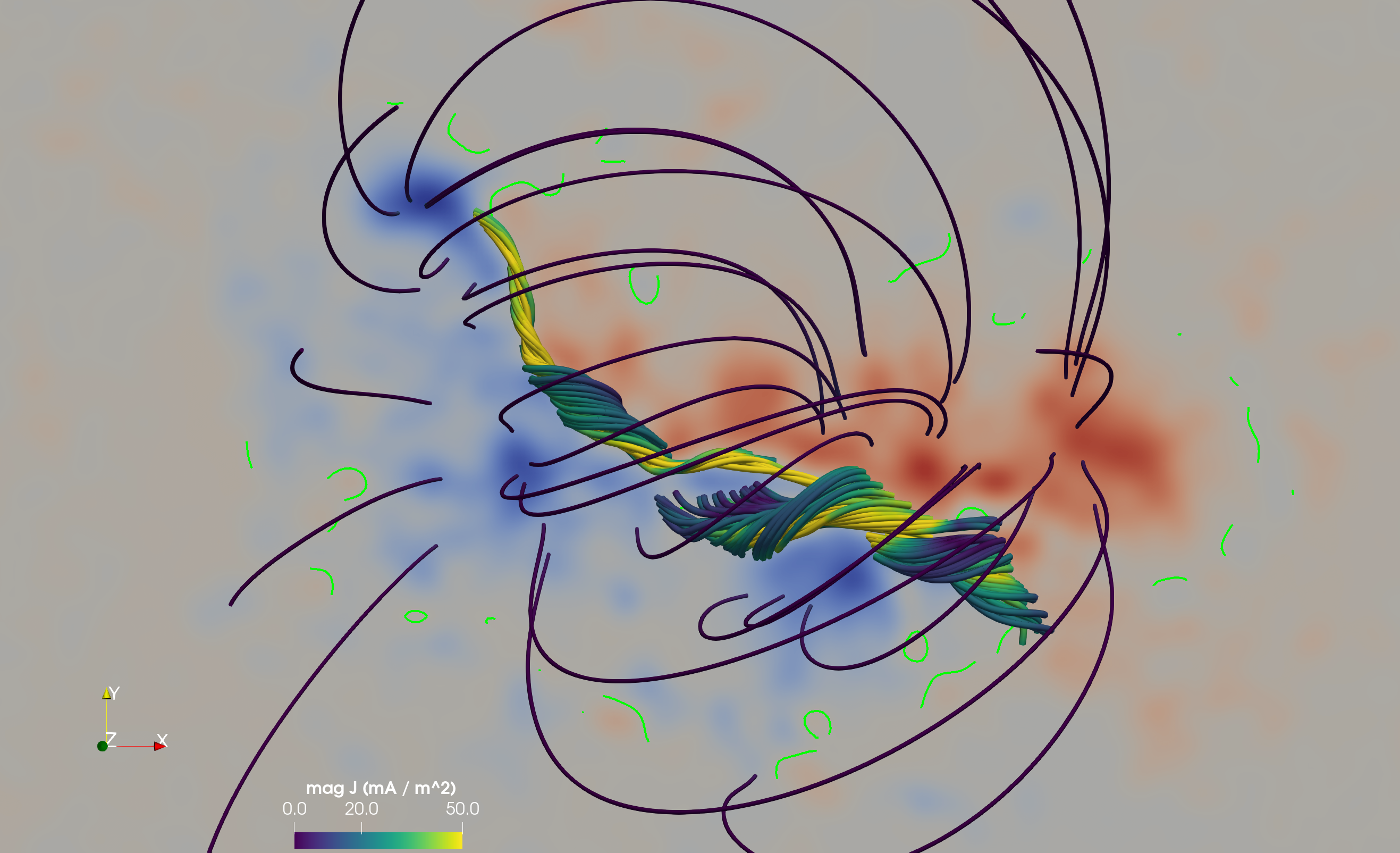}
            \vspace{0.2cm}
        \caption{The NLFFF model for AR 11429 on 2012 March 6 23:36 UT using the optimization method. The background color denotes the $B_z$ component at the photosphere. The green $B_z=0$ contours indicate the main PILs. The field lines are colored by the magnitude of the current density $|\mbf{J}|$, sampled along the line.}
        \label{fig:field_lines_AR11429_opti}
    \end{figure*}

    The optimization solution for AR 11429 has a normalized energy of $E/E_0 = 1.13$, which is less than that of the CFIT $N$ solution, \added{but similar to the CFIT $P$ solution \citep[see, e.g.,][for a discussion on the variation of energies in different NLFFF methods]{derosa_influence_2015}}. Selected field lines for the optimization solution are shown in \autoref{fig:field_lines_AR11429_opti}. A low-lying flux rope extends along both the upper-left and lower-right half of the PIL as indicated by the yellow field lines. The flux rope is more consistent with that found by \citet{chintzoglou2015_ar11429}, where a similar method was used.        

\begin{figure*}
        \centering
        \plotone{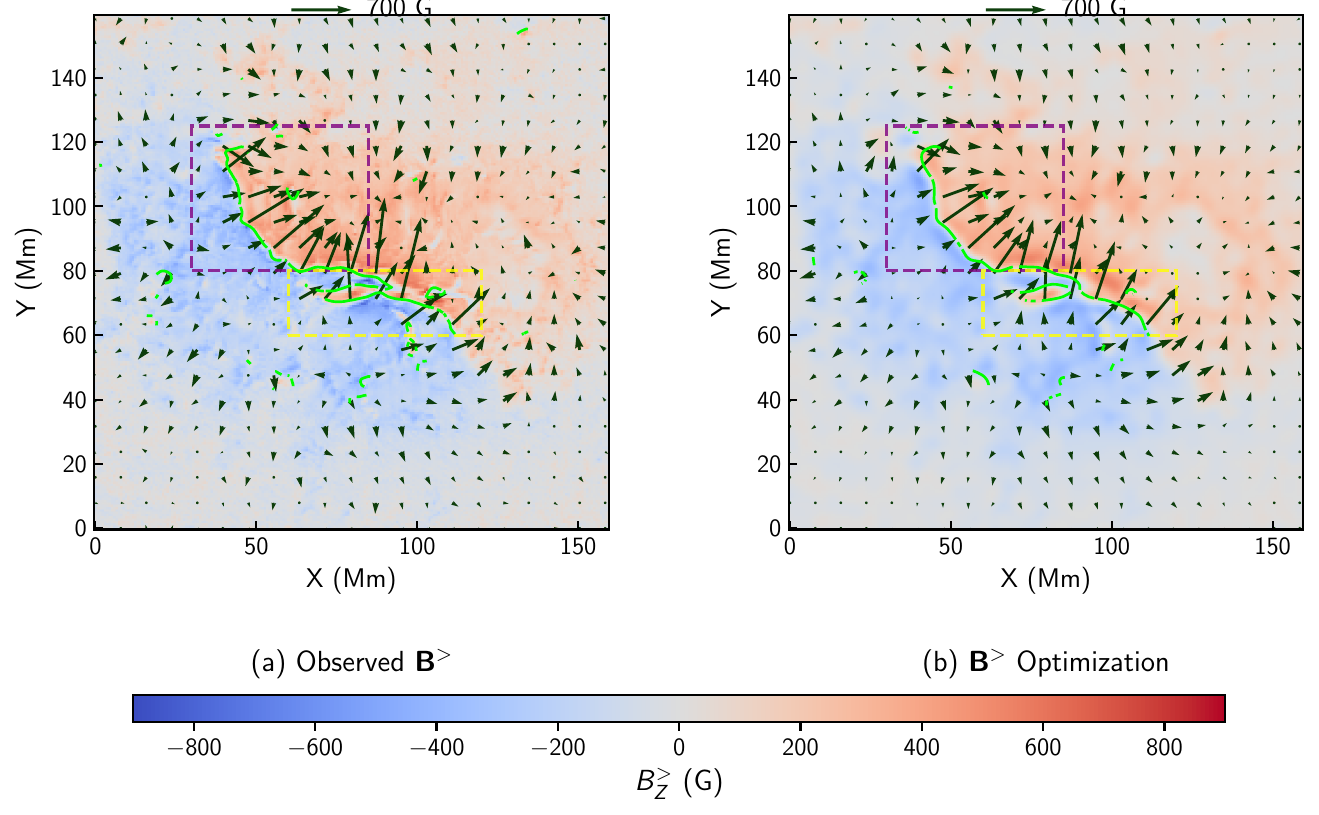}
        \caption{The photospheric signatures of coronal currents, ${\textbf{B}^>(x,y)}$ derived from (a) SHARP photospheric vector magnetogram, and (b) lower boundary data of the optimization solution for AR 11429 on 2012 March 6 23:36 UT. The background color map in each panel denotes $B_z^>(x,y)$, the black contours denote PILs corresponding to $B_z(x,y) = 0$, and the dark green vectors denote the horizontal component $\textbf{B}_h^>(x, y)$. The color and arrow scales are the same for both panels.}
        \label{fig:AR11429_B>_Opti_comparison}
    \end{figure*}

\begin{figure*}
        \centering
        \plotone{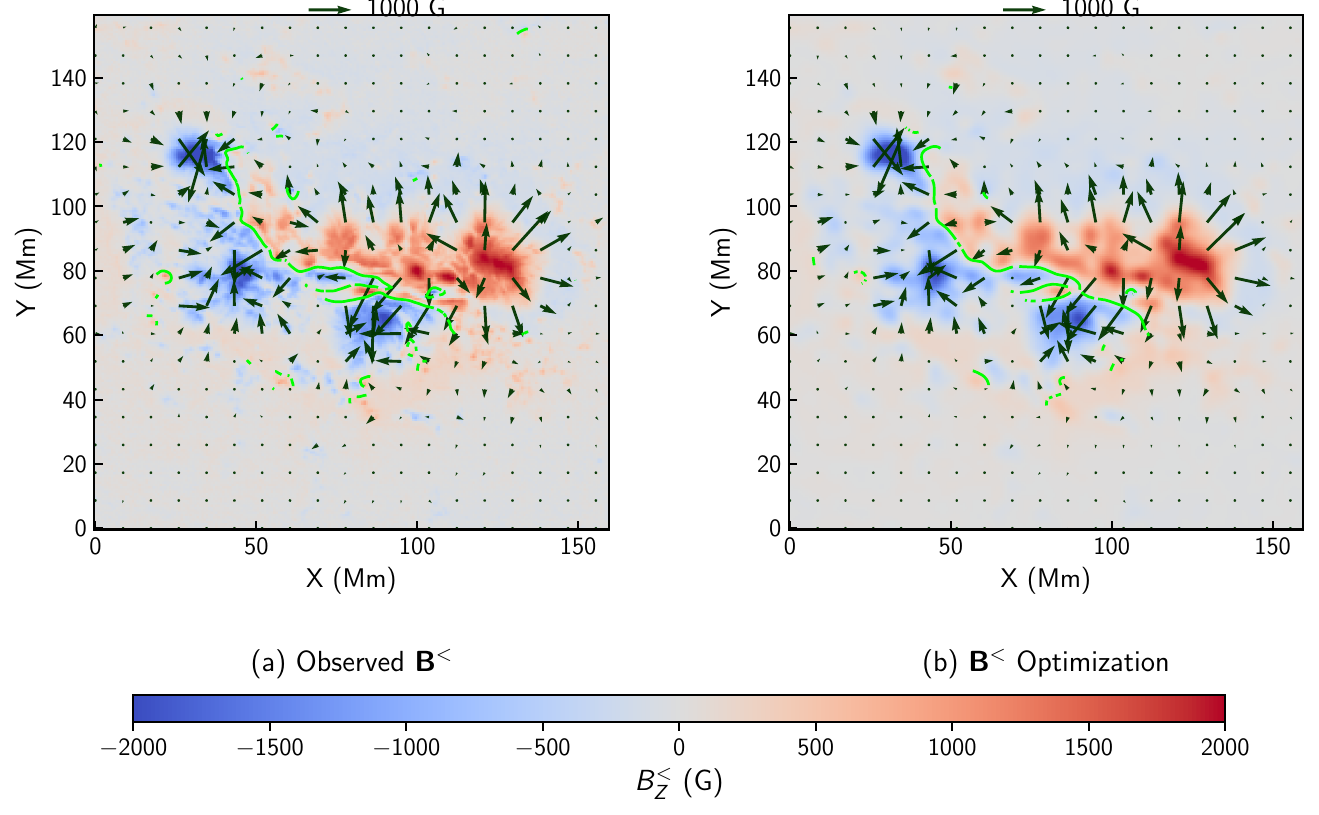}
        \caption{The photospheric signatures of subphotospheric currents, $\blt$ derived from (a) the SHARP photospheric vector magnetogram, and (b) the lower boundary data of the optimization solution for AR 11429 on 2012 March 6 23:36 UT. The background color map in each panel denotes $B_z^<(x,y)$, the black contours show the main PILs corresponding to $B_z(x,y) = 0$, and the purple vectors show the horizontal component $\textbf{B}_h^<(x, y)$. The color and arrow scales are the same for both panels.}
        \label{fig:AR11429_B<_Opti_comparison}
    \end{figure*}

\begin{figure*}
        \centering
        \plotone{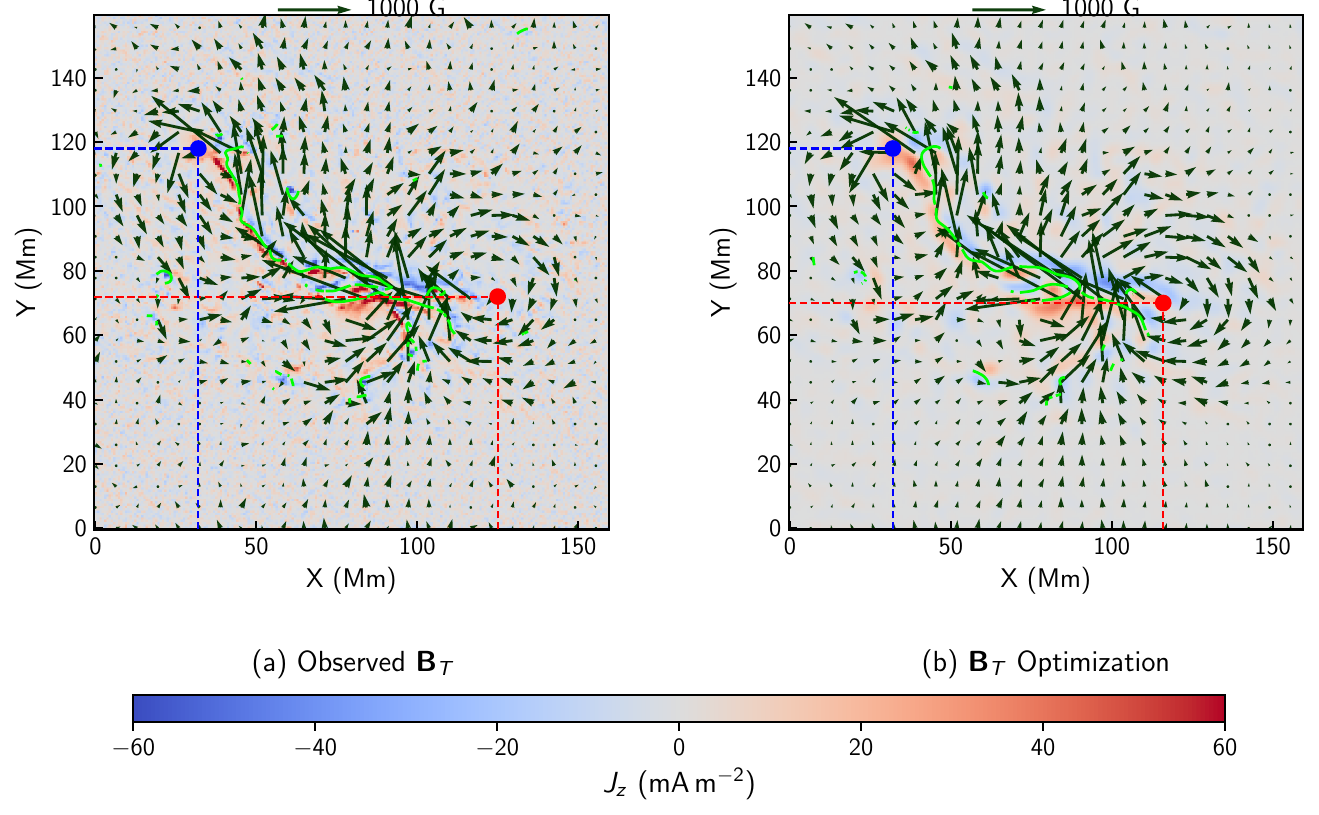}
        \caption{The toroidal field, ${\textbf{B}_{T}(x,y)}$,  produced by vertical currents, $J_z(x,y)$, derived from (a) SHARP photospheric vector magnetogram, and (b) lower boundary data of the optimization solution for AR 11429 on 2012 March 6 23:36 UT. The background color map in each panel denotes $J_z(x,y)$, the green contours denote PILs corresponding to $B_z(x,y)=0$, and the orange vectors denote $\btor$.}
        \label{fig:AR11429_B_T_opti_comparison}
    \end{figure*}

    Figures \ref{fig:AR11429_B>_Opti_comparison}-\ref{fig:AR11429_B_T_opti_comparison} show the result of applying Gaussian separation to the SHARP photospheric vector magnetogram and the lower boundary data of the optimization solution for AR 11429. The $\mathbf{B}_h^>(x,y,0)$ vectors in panel (b) of \autoref{fig:AR11429_B>_Opti_comparison} are qualitatively similar to the those in panel (a) along the entire length of the PIL. The $B_z^>(x,y,0)$ map in panel (b) of \autoref{fig:AR11429_B>_Opti_comparison} is also qualitatively similar to that in panel (b), but is smoother due to the preprocessing of the boundary data. 

    \added{\autoref{fig:AR11429_B_T_opti_comparison} compares $\btor$ derived from the SHARP vector magnetogram (panel (a)) and the lower boundary of the optimization solution (panel (b)), similar to \autoref{fig:AR11429_B_T_comparison}. The regions where  vertical currents are directed out of (blue circle) and into  (red circle) the photosphere are qualitatively similar in both panels. The sheared $\btor$ field in the photosphere along the main PIL is also reproduced accurately in the optimization solution.}

    The use of preprocessing the lower boundary magnetic field data ensures that the optimization solution does not match the observed $\mathbf{B}_h(x,y,0)$ or $B_z(x,y,0)$ at the photosphere. Denoting the difference between the spectral functions from the vector magnetogram and the optimization solution as $\Delta \tilde{\chi}^{\lessgtr}_{\mathrm{OPTI}} = \tilde{\chi}^{\lessgtr}_{\mathrm{obs}} - \tilde{\chi}^{\lessgtr}_{\mathrm{OPTI}}$, Equation \eqref{eq:theoretical_diff_b_gt_lt}, with $\tilde{\chi}^{\lessgtr}_{\mathrm{CFIT}}$ replaced by $\tilde{\chi}^{\lessgtr}_{\mathrm{OPTI}}$, implies that differences in $\bgt$ and $\blt$ between the vector magnetogram and the optimization solution are due to differences in both $\mathbf{B}_h(x,y,0)$ and $B_z(x,y,0)$. In addition, the magnitude of the differences in $\bgt$ and $\blt$ may be different, as discussed below. 

    \added{The rms values of $|\Delta \mathbf{B}_h|$ and $|\Delta B_z|$ for the total field and the Gaussian separated components are listed in \autoref{tab:diff_bgt_blt_CFIT}. The rms differences in $\bgt$ and $\blt$ in the optimization solution are less than those for the CFIT $N$ solution by about $50\%$ and $35\%$ respectively. Although the CFIT method preserves $B_z(x,y, 0)$ at the lower boundary, $|\Delta \mathbf{B}_h(x,y, 0)|$ in the CFIT solution is large compared to the optimization solution. Because $\bgt$ and $\blt$ are calculated using all three components of the field in the lower boundary, the optimization method has $\bgt$ and $\blt$ values that are closer to those derived from the vector magnetogram.}   

\begin{figure*}
        \centering
        \plotone{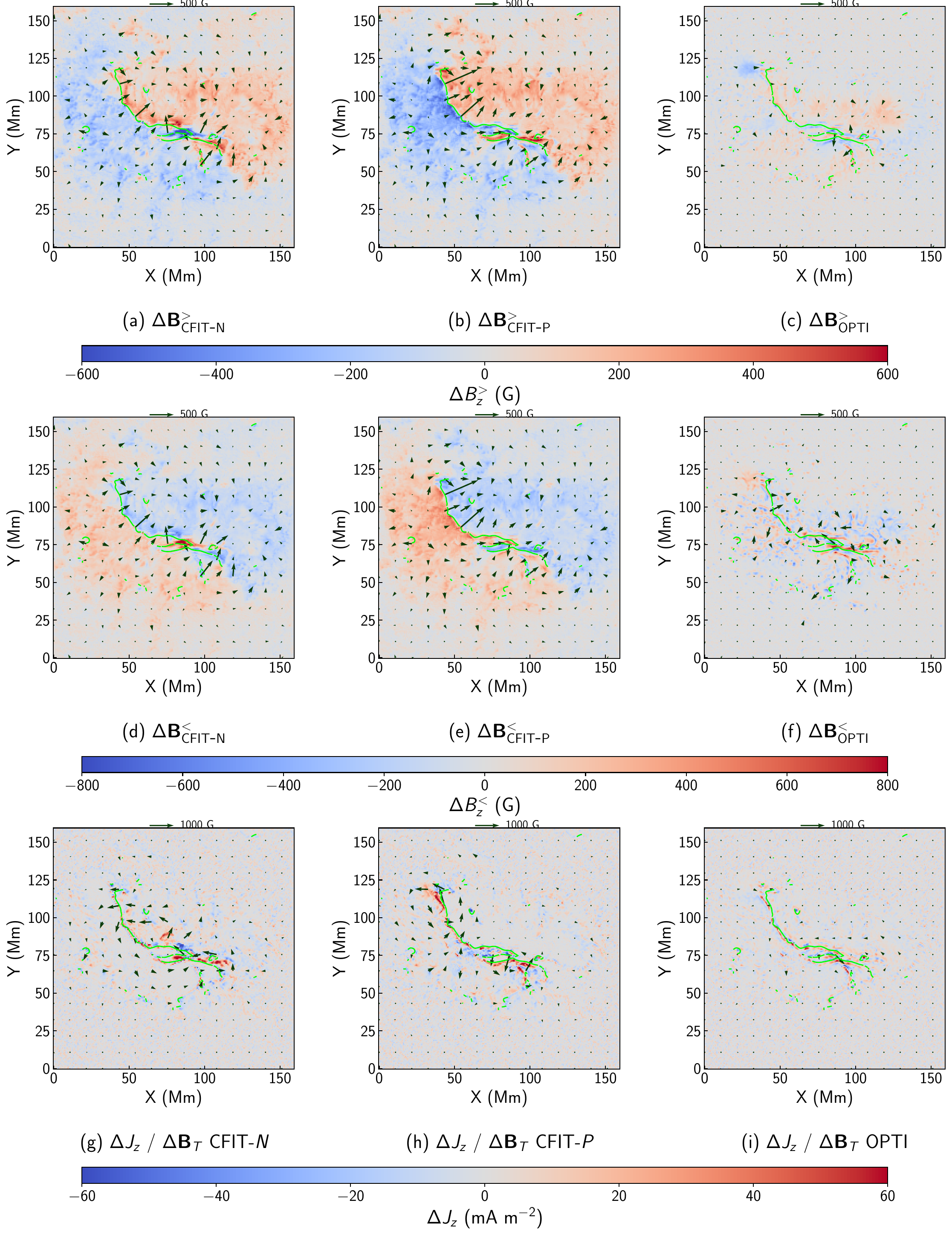}
        \caption{Spatial distribution of the differences between the Gaussian separation components derived from the vector magnetogram and the lower boundary of the models. The background color map in panels (a) to (c) indicate $\Delta B_z^>$ for each of the models, and the vectors indicate $\Delta \mbf{B}_h^>$. Panels (d) to (f) show similar plots for $\Delta B_z^<$ and $\Delta \mathbf{B}_h^<$. The background color map in panels (g) to (i) shows $\Delta J_z$, with the vectors indicating $\Delta \mathbf{B}_T$. The green contours in each panel indicate the main PIL.}
        \label{fig:AR11429_gs_jz_diff_opti_cfit}
    \end{figure*}

\added{\autoref{fig:AR11429_gs_jz_diff_opti_cfit} illustrates the spatial distribution of the differences in the Gaussian separation components derived from the vector magnetogram and the lower boundary of all the models discussed.  Panels (a) and (b) show that $\Delta B_z^>$ is positive in the positive polarity, and negative in the negative polarity, indicating that the magnitude of $B_z^>$ in both the CFIT solutions is reduced compared to the magnetogram. Correspondingly, panels (d) and (e) show that the magnitude of $B_z^<$ is increased in both the CFIT solutions. These differences, together with $\Delta J_z$ (panels (g) and (h)), are significantly larger near the main PIL. In addition, the differences in $J_z$ are larger in the positive polarity for the $N$ solution, and in the negative polarity for the $P$ solution. This is expected, since the $\alpha$ values that are mapped to the opposite polarity do not necessarily match the vector magnetogram values of $\alpha$. The differences in all three Gaussian separation components for the optimization solution (panels (c), (f), and (i)) are significantly less compared to the CFIT solutions, indicating a lower boundary more consistent with the observed magnetogram.} 

\begin{figure*}
    \plottwo{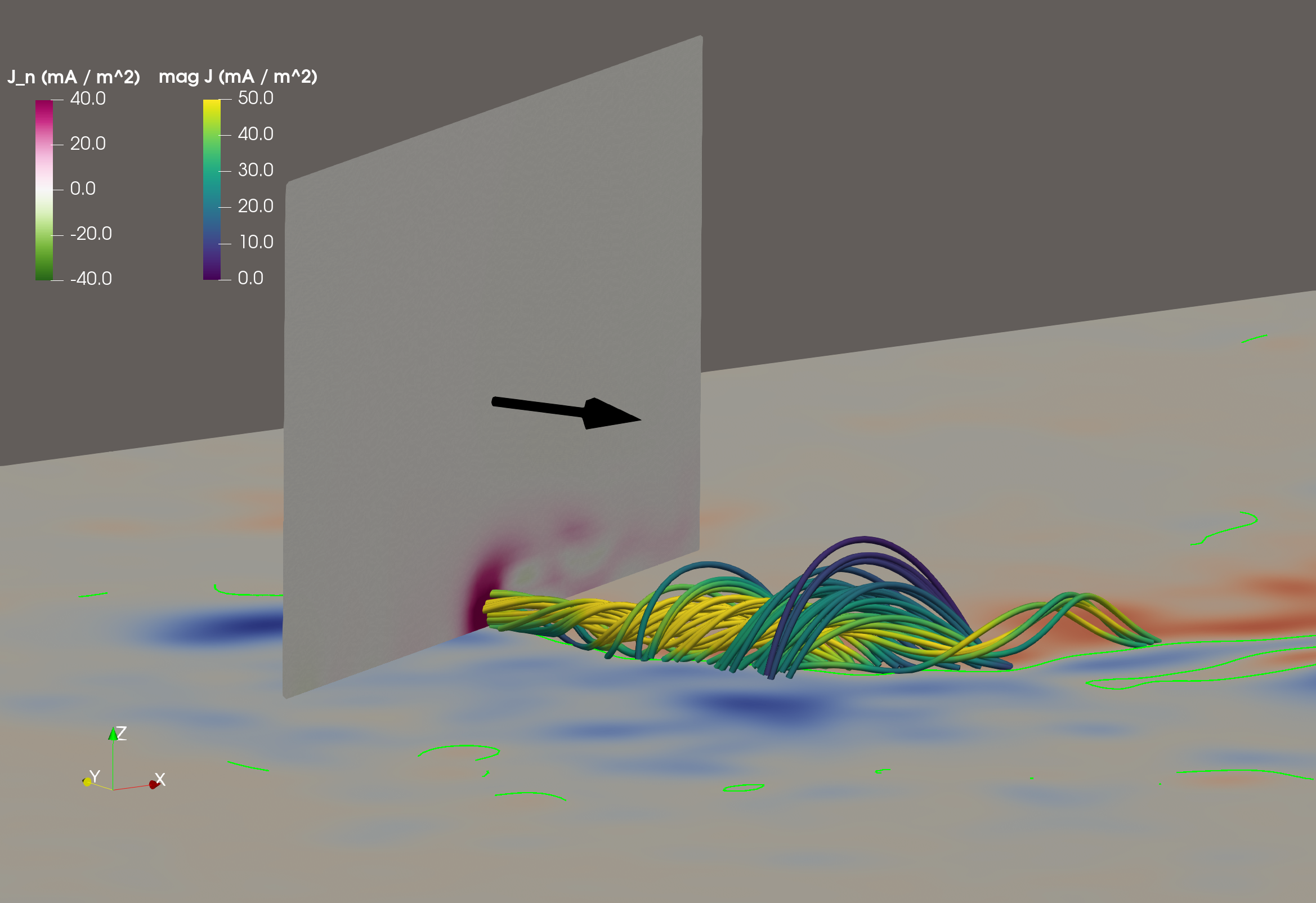}{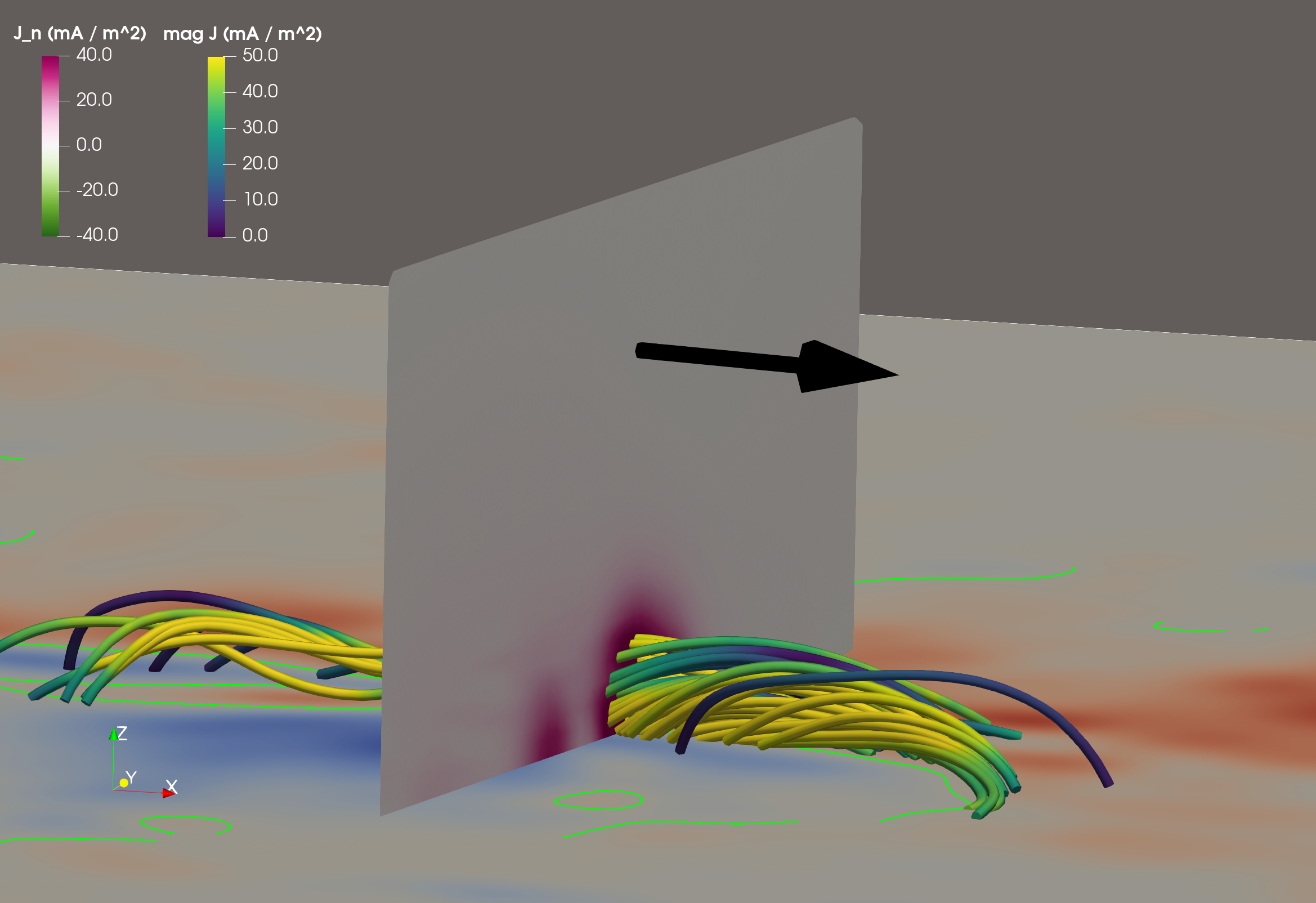}
    \caption{Flux rope and associated current density distribution for AR 11429, derived from the optimization solution. The vertical slices considered are the same as those shown in panel (a) of \autoref{fig:AR11429_Upper_Lower_PIL_CFIT}. The left panel shows the flux rope and current density in the upper section of the PIL, and the right panel shows the bottom section of the PIL. The field lines are colored by $|\mbf{J}_{\mathrm{OPTI}}(x,y, z)|$ sampled along the line. The normal current density in the vertical slice perpendicular to the PIL, $J_n$ is shown. Positive values of $J_n$ indicate currents flowing in the direction of the normal vector to the plane. The lower boundary shows $B_z(x,y)$ and the main PIL, similar to previous figures.}
    \label{fig:AR11429_Upper_Lower_PIL_Opti}
\end{figure*}

\added{We now consider the currents $\mathbf{J}_{\mathrm{OPTI}}(x,y,z)$ in the optimization solution and perform an analysis similar to Section \ref{sec:cfit_results}. \autoref{fig:AR11429_Upper_Lower_PIL_Opti} shows selected field lines in the upper-left (left panel) and lower-right (right panel) sections of the PIL. The field lines are colored by the magnitude of the current density sampled along the line. The structure of the flux rope along the upper-left section of the PIL is qualitatively similar to the one seen in the CFIT $N$ solution (see left panel of \autoref{fig:AR11429_Upper_Lower_PIL_CFIT}), but is closer to the photosphere. The normal current density distribution in the same vertical cross-section as in \autoref{fig:AR11429_Upper_Lower_PIL_CFIT} is also qualitatively similar. The peak value of the current density is $|J_n| = 53.3 \, \mathrm{mA} \, \mathrm{m}^{-2}$, at around $z=1.44$ Mm. Despite the differences in the flux rope and current density between the optimization and CFIT $N$ solutions, the distribution of $\bgt$ in both methods is qualitatively similar, in this section of the PIL.}

\added{The right panel of \autoref{fig:AR11429_Upper_Lower_PIL_Opti} shows that the field lines in the optimization solution are sheared along the lower-right section of the PIL, unlike the field lines of the CFIT solution, which are more nearly perpendicular to the PIL (see right panel of \autoref{fig:AR11429_Upper_Lower_PIL_CFIT}). The peak value of the current density in the vertical cross-section is $|J_n| = 55.17 \, \mathrm{mA} \, \mathrm{m}^{-2}$, at around $z=2.16$ Mm. Since the structure of $\bgt$ derived from the optimization solution is more similar to that derived from the vector magnetogram, the field lines in the optimization solution may be modeled more accurately compared to the CFIT solution.}

\added{Similar to Section \ref{sec:cfit_results}, we assess the magnitude of the coronal currents in the optimization solution by computing $I_C$ and $I_S$ as defined in Equations \eqref{eq:amp_law} and \eqref{eq:ic_decomp}. \autoref{tab:integrals_opti} lists the integrals evaluated using $\mbf{B}_{\rm{OPTI}}(x,y,z)$ and $\mbf{J}_{\rm{OPTI}}(x,y,z)$, along with $I_1$ evaluated using $\mbf{B}_{\rm{obs}}(x,y,0)$, for the vertical cross-sections indicated in panel (a) of \autoref{fig:iphot_icor_cfit}.}
\begin{table}[ht]
    \centering
    \caption{Ampère's law integrals
    for the vertical cross-sections of the optimization solution.}
    \begin{tabular}{lcc}
        \toprule
        Quantity  & Slice 1 ($10^9$\,A) & Slice 2 ($10^9$\,A) \\
        \midrule
        $I_1$ (observed)   & 1915.54  & 3076.77 \\
        $I_1$ (OPTI)       & 2199.24  & 3247.33 \\
        $I_2$              &  -34.01  &   46.79 \\
        $I_3$              &  41.47 &     72.77 \\
        $I_4$              &   18.39  &   84.19 \\
        \midrule
        $I_C$       & 2225.09  & 3451.08 \\
        $I_S$ & 2123.31  & 3372.39 \\
        \bottomrule
    \end{tabular}
    \label{tab:integrals_opti}
\end{table}

\added{In both slices, $I_C \approx I_S$, as expected since the optimization method is also consistent with Ampère's law, similar to the CFIT method. The value of $I_1$ evaluated using $\mbf{B}_{\rm{obs}}(x,y,0)$ in the upper section of the PIL (Slice 1 in panel (a) of \autoref{fig:iphot_icor_cfit}) is around 15\% less than that evaluated using $\mbf{B}_{\rm{OPTI}}(x,y,0)$, and around 5\% less in the lower section of the PIL (Slice 2 in panel (a) of \autoref{fig:iphot_icor_cfit}). The discrepancies in $I_1$ for the optimization solution are smaller than those for the CFIT $N$ solution, suggesting that the coronal currents in the optimization solution may be modeled more accurately.} 

\section{Discussion and Conclusions} \label{sec:discussion}
    NLFFF models have been widely used to represent the coronal magnetic field on the Sun. A key question is whether the models accurately represent the coronal electric currents, which are necessary to study the mechanisms of energetic events that take place on the Sun. Recent studies by \citet{schuck_origin_2022} and \citet{welsch_photospheric_2022} demonstrated that the imprints of coronal currents can be obtained from photospheric vector magnetograms of active regions, using Gaussian separation. In this work, we have applied Gaussian separation to NLFFF models of AR 11429 on 2012 March 6 23:36 UT, constructed using the optimization and the CFIT methods. We then compared the photospheric imprints of the coronal currents in the models with those derived from the vector magnetogram, which is the input for the models. 
    
    The structure of $\mathbf{B}_h^>(x,y, 0)$ along the main PIL of AR 11429, derived from the vector magnetogram, implies a coronal current flowing above and parallel to the PIL, similar to AR 12673 and AR 11158 analyzed by \citet{schuck_origin_2022} and \citet{welsch_photospheric_2022} respectively. We find that the sheared field along the PIL is primarily due to $\btor$, which is produced by $J_z(x,y, 0)$. The structure of $\bgt$ and $\btor$ in the models is qualitatively similar to that of the vector magnetogram, but there are significant quantitative differences. Using the field lines in the models, together with their associated current density distributions, we identify the flux rope along the main PIL that produces the observed structure of $\mathbf{B}_h^>(x,y, 0)$. The Gaussian separated components in the optimization model match those of the vector magnetogram better than the CFIT model. The differences in the Gaussian separated components of the vector magnetogram and the models are due to differences in the lower boundary data. As previous studies \citep[e.g.,][]{schrijver2006, metcalf2008, derosa2009, derosa_influence_2015} have shown, the changes made to the magnetic field at the lower boundary to achieve a force-free solution result in the NLFFF models having boundary data which are different from the input vector magnetogram data. However, despite the limitations of the models, the NLFFFs are able to qualitatively reproduce the signatures of key coronal current structures. 

    While the analysis of the Gaussian separated components validates the signatures of the coronal currents in the models, it is also necessary to verify whether the NLFFF solutions are force- and divergence-free in the coronal volume ($z>0$). There is a trade-off between the force-freeness of the solution (and, in the optimization case, the divergence-freeness) and its consistency with the observed photospheric field. In the optimization method, preprocessing is necessary to achieve a more force-free solution. If no preprocessing were applied, the magnetic field at the lower boundary of the solution would exactly match the observed photospheric field, but the solution may not be force- or divergence-free. However, the photospheric signatures of the coronal currents in the preprocessed boundary will only be consistent with the currents in the volume if the solution is force- and divergence-free. We assess the degree to which the NLFFF solutions in this work are divergence- and force-free using the metrics listed in \autoref{tab:nlfff_metrics}. We measure the divergence-freeness using the modified fractional flux, $\left<|f_d|\right> = \left< |\nabla \cdot \mbf{B}|_i / |\mathbf{B}|_i \right>$ (where $i$ denotes each voxel in the grid) \citep[see, e.g., ][]{gilchrist_divB}, where zero indicates a perfectly divergence-free field. We measure the force-freeness using the current-weighted mean sine of the angle between $\mbf{J}$ and $\mbf{B}$, $\left<\textrm{CW} \sin\theta \right>$, where zero indicates a perfectly force-free field \citep[see, e.g.,][]{wheat_opti}. \added{The metrics for both the CFIT solutions and the optimization solution in this work are comparable to values in the literature \citep[see, e.g., ][]{thalmann2022}, indicating that the solutions are sufficiently force- and divergence-free in the coronal volume. The larger values of $\left<\textrm{CW} \sin\theta \right>$ for the solutions in this work may be because a significant portion of the magnetic field in the volume is close to potential.}   
    \begin{table}[ht]
    \centering
    \caption{NLFFF metrics for AR 11429}
    \begin{tabular}{ccc}
        \toprule
         Solution & $\left<|f_d|\right>$ ($\times 10^{-9} \, \textrm{m}^{-1}$) & $\left<\textrm{CW} \sin\theta \right>$ \\ 
         \midrule \midrule 
         CFIT $N$ & 5.03 & 0.21 \\ 
         \midrule
         CFIT $P$ & 4.89 & 0.24 \\
         \midrule
         Optimization & 8.36 & 0.32 \\
        \bottomrule
    \end{tabular}

    \label{tab:nlfff_metrics}
\end{table}

    The analysis in this work has been carried out for a single active region at a particular time. A more detailed study of multiple active regions, over the course of their evolution is warranted. Previous studies have found flare-associated changes in the horizontal and vertical components of the photospheric magnetic field \citep[see, e.g.,][]{petrie2012abrupt, sun2017flarechanges, duran2018flarechanges, liu2018evolution, yadav2023statistical}, and electric currents \citep[e.g.,][]{janvier2016evolution, barczynski2020electric}. Similar changes in $\bgt$, $\blt$ and $\btor$ during flares, if any, should be investigated. 

    It is also of interest to improve models of the coronal magnetic field by using information inferred from Gaussian separation. One approach for the optimization method is to include an additional term in the functional defined in Equation \eqref{eq:opti_functional}, similar to Equation \eqref{eq:opti_add_functional}, which minimizes differences between $\bgt$ derived from the vector magnetogram and the lower boundary of the optimization solution. Recently, \citet{jarolim2023pinn} introduced a method to determine the coronal magnetic field using physics-informed neural networks (PINNs). In the PINN approach, the coronal magnetic field is represented as a neural network trained to minimize losses associated with deviations from the observed vector magnetogram at the lower boundary and the force-free conditions \citep[see][for details]{jarolim2023pinn}. An additional loss term could be added to constrain the model to match the $\bgt$ derived from the vector magnetogram, although this would need to be incorporated during the training of the neural network.

    Gaussian separation can also be applied to other models such as magnetofrictional and MHD models, to validate the accuracy of the reproduced coronal currents. The locations of currents in the MHD model which contribute to the Gaussian separated components and how they evolve over time could be investigated. 
    
    \added{A more detailed comparison of the MBSL \citep[][]{titov2025} and Gaussian decomposition methods, applied to both vector magnetograms and coronal magnetic field models is also warranted. The Gaussian separation provides a unique and well-defined decomposition of the photospheric field based on the physical separation of currents, while the MBSL approach provides a method to insert coronal currents into a potential field, while preserving the observed normal component of the field at the lower boundary. While each decomposition may be expressed as linear combinations of the other, it is not clear if the photospheric signatures of coronal currents would be similar.}  
    
\section*{Acknowledgments}
    AGI and MSW acknowledge support from the Australian Research Council via the Discovery Project grant DP230101240. BTW and SAG acknowledge salary support from the U.S. National Science Foundation via award AGS 2302697, which also supported travel costs for AGI and MSW to attend the SHINE Workshop 2025 in Charleston. NASA's SDO satellite and the HMI instrument were joint efforts by many teams and individuals, whose efforts to produce the HMI magnetogram that we analyzed here are greatly appreciated. The Python implementation of the Gaussian separation method was developed by SAG with support from NSF award 2302698. Any opinions, findings, conclusions, or recommendations expressed in this material are those of the authors and do not necessarily reflect the views of the National Science Foundation or the Australian Research Council. 

\appendix
\section{The field due to a ring current}
\label{sec:appendix}
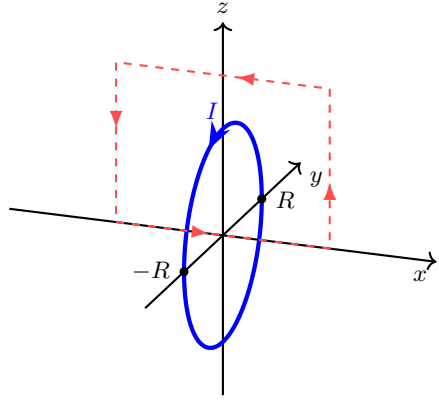
\begin{figure}[ht!]
\centering
\tdplotsetmaincoords{70}{20}
\begin{tikzpicture}[tdplot_main_coords, scale=1.5]
    \draw[thick, ->] (-2.0,0,0) -- (2.0,0,0) node[anchor=north east]{$x$};
    \draw[thick, ->] (0,-2,0) -- (0,2,0) node[anchor=north west]{$y$};
    \draw[thick, ->] (0,0,-1.5) -- (0,0,2.0) node[anchor=south]{$z$};

    \draw[ultra thick, blue, 
        decoration={markings, mark=at position 0.25 with {\arrow{Stealth}}}, 
        postaction={decorate}] 
        plot[domain=0:360, samples=100, variable=\t] (0, {cos(\t)}, {sin(\t)});
    
    \node[blue, anchor=south] at (-0.1, 0, 1) {$I$};

    \filldraw[black] (0, 1, 0) circle (1.0pt); 
    \node[black, anchor=west] at (0.05, 1, 0) {$R$};
    
    \filldraw[black] (0, -1, 0) circle (1.0pt); 
    \node[black, anchor=east] at (-0.05, -1, 0) {$-R$};

    \coordinate (A) at (-1.0, 0, 0);   
    \coordinate (B) at (1.0, 0, 0);    
    \coordinate (C) at (1.0, 0, 1.5);  
    \coordinate (D) at (-1.0, 0, 1.5); 

    \draw[thick, dashed, red!70, 
        decoration={markings, mark=at position 0.125 with {\arrow{Latex}},
                              mark=at position 0.375 with {\arrow{Latex}},
                              mark=at position 0.625 with {\arrow{Latex}},
                              mark=at position 0.875 with {\arrow{Latex}}}, 
        postaction={decorate}] 
        (A) -- (B) -- (C) -- (D) -- cycle;
\end{tikzpicture}
\caption{A ring current in the $y-z$ plane, with total current $I$ and radius $R$. The red dashed line shows the Amperian loop used to calculate the net current above the $x-y$ plane.} 
\label{fig:ring_current}
\end{figure}

To illustrate the contribution of the components in the Gaussian separation to the line integral of the total field $\mbf{B}(x,y,0)$, we consider a ring current, as shown in Figure~\ref{fig:ring_current}. The ring current has the $x$-axis as the axis of symmetry, radius $R$, and total current $I$. The figure shows (in red) a rectangular Amperian circuit with one side along the $x$-axis. If the rectangle is expanded to infinity, then the only contribution will be from the side along the $x$-axis. 

The current density of the ring current in the $x-y$ plane is
\begin{equation}
J_z(x,y,0)= I\left[\delta (\mbf{r}-R\widehat{\mbf{y}})-\delta (\mbf{r}+R\widehat{\mbf{y}})\right]
\label{eq:jz_ring}
\end{equation}
where $\mbf{r}=x\widehat{\mbf{x}}+y\widehat{\mbf{y}}$. 

\begin{figure}[!ht]
    \centering
    \includegraphics[width=0.5\linewidth]{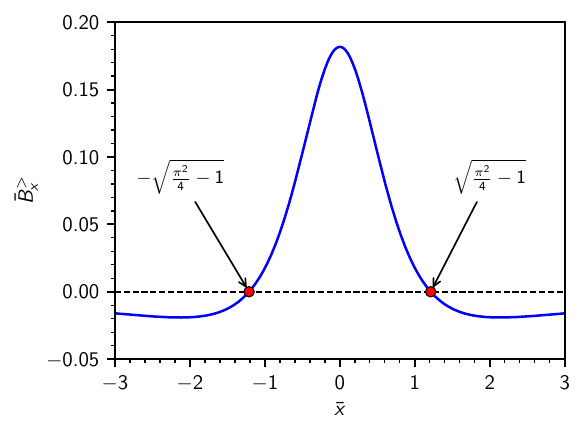}
    \caption{The functional form of $B_x^{>}(x)$ in non-dimensional units. Equation \eqref{eq:ring_current_bx_gt_eq} is non-dimensionalized by choosing $\bar{x} = x/R$ and $\bar{B}_x^> = B_x^>/B_0$, where $B_0 = \mu_0 I/2R$. The points at $|\bar{x}|=\sqrt{\pi^2/4 -1}$ indicate where the field $\bar{B}_x^>$ becomes negative.
    \label{fig:ring_current_bx_gt_plot}}
\end{figure}

The field $\btor$ is defined by the toroidal scalar potential $T(x,y,0)$ according to Equations~(\ref{eq:btor})-(\ref{eq:tor_pot}), so we need to solve
\begin{equation}
\nabla_h^2T(x,y,0)= -\mu_0 J_z(x,y, 0).
\label{eq:tor_pot2}
\end{equation}
The solution (in 2-D) to $\nabla^2 G(\mbf{r},\mbf{r}_0) = \delta (\mbf{r}-\mbf{r}_0)$ is the Green's function $G(\mbf{r},\mbf{r}_0) = \frac{1}{2\pi}\ln |\mbf{x}-\mbf{x}_0|.$ Hence it follows that the solution to Equation~(\ref{eq:tor_pot2}) with the current distribution given by Equation~(\ref{eq:jz_ring}) is
\begin{equation}
T(x,y) = \frac{\mu_0 I}{2\pi}\left(\ln\sqrt{x^2+(y+R)^2}-\ln\sqrt{x^2+(y-R)^2}\right).
\end{equation}
The components of $\btor$ are then obtained using Equation~(\ref{eq:btor}):
\begin{equation}
B_x^T(x,y) = \frac{\mu_0 I}{2\pi}\left(\frac{y+R}{x^2+(y+R)^2}-\frac{y-R}{x^2+(y-R)^2}\right)
\label{eq:btor_x_ring}
\end{equation}
and
\begin{equation}
B_y^T(x,y) = \frac{\mu_0 I}{2\pi}x\left(\frac{1}{x^2+(y+R)^2}-\frac{1}{x^2+(y-R)^2}\right).
\end{equation}

We consider the integral of $B_x^T(x,y)$ along a line parallel to the $x$ axis. Using Equation~(\ref{eq:btor_x_ring}) it follows that
\begin{equation}
\int_{-\infty}^{\infty}B_x^T(x,y)\,dx = 
\lim_{X\rightarrow\infty}\frac{\mu_0 I}{2\pi}
\left( \left[\arctan\frac{x}{y+R}\right]_{-X}^{X}-\left[\arctan\frac{x}{y-R}\right]_{-X}^{X}\right)
\end{equation}
which evaluates to
\begin{equation}
\int_{-\infty}^{\infty}B_x^T(x,y)\,dx = \left\{
    \begin{array}{ll}
      \mu_0 I, & \mbox{if $|y|<R$}\\
      0, & \mbox{if $|y|>R$}.
    \end{array}
  \right.
\end{equation}
Hence if the Amperian circuit intersects the ring current, the integral evaluates to the total expected from Ampère's law, and if the circuit is outside the ring current the integral evaluates to zero. This implies that if we integrate the total field $B_x(x,y,0) = B_x^T(x,y)+B_x^{>}(x,y,0)+B_x^{<}(x,y,0)=B_x^T(x,y)+2B_x^{>}(x,y,0)$ (by symmetry), then the contribution from $B_x^{>}(x,y,0)$ will be zero.

We can show the explicit contributions from each component in the Gaussian separation when the Amperian integral is along the $x$-axis, as illustrated in Figure~\ref{fig:ring_current}. The total field along the axis of the ring current is \citep[e.g.,][]{Griffiths1989}
\begin{equation}
B_x(x) = \frac{\mu_0 I}{2}\frac{R^2}{\left(x^2+R^2\right)^{3/2}}.
\end{equation}
Using $B_x^{>}=\frac{1}{2}(B_x-B_x^T)$ together with Equation~(\ref{eq:btor_x_ring}) with $y=0$ gives
\begin{equation}
B_x^{>}(x) = \frac{\mu_0 IR}{2}\left(\frac{1}{2}\frac{R}{\left(x^2+R^2\right)^{3/2}}-\frac{1}{\pi}\frac{1}{\left(x^2+R^2\right)}\right).
\label{eq:ring_current_bx_gt_eq}
\end{equation}
It is straightforward to show that $\int_{-\infty}^{\infty}B_x^{>}(x)\,dx = 0$. An interesting point here is that the function $B_x^{>}(x)$ becomes negative sufficiently far away from the ring current. Specifically, $B_x^{>}(x)<0$ when $|x|>R\sqrt{\pi^2/4-1}$. This is counter-intuitive because the right hand rule leads us to expect that the contribution from the current in the corona will always be in the positive-$x$ direction. Figure~\ref{fig:ring_current_bx_gt_plot} shows the functional form of $B_x^{>}(x)$. 

\bibliography{bibliography}{}
\bibliographystyle{aasjournalv7}
\end{document}